\documentclass[superscriptaddress,showpacs,aps,twocolumn,prb,floatfix,longbibliography]{revtex4-1}
\usepackage{bm,amsmath,amssymb}
\usepackage{amssymb}
\usepackage{amsmath}
\usepackage{cancel}
\usepackage{commath}
\usepackage{graphicx,bm}
\usepackage{verbatim}
\usepackage[latin1]{inputenc}
\usepackage{xcolor}
\usepackage[a4paper]{hyperref}
\hypersetup{colorlinks=true,linktoc=all,linkcolor=blue,breaklinks=true,citecolor=blue,urlcolor=blue}

\newcommand{\dmu}{\Delta\mu}
\newcommand{\muc}{\mu_\text{c}}
\newcommand{\muh}{\mu_\text{h}}

\newcommand{\dT}{\Delta T}
\newcommand{\Tc}{T_\text{c}}
\newcommand{\Th}{T_\text{h}}

\newcommand{\df}{\Delta f(\varepsilon)}

\newcommand{\TQSH}{\mathcal{T}^\text{QSH}}
\newcommand{\TRB}{\mathcal{T}^\text{RB}}
\newcommand{\Twell}{\mathcal{T}^\text{well}}
\newcommand{\Tstep}{\mathcal{T}^\text{step}}
\newcommand{\Tpeak}{\mathcal{T}^\text{Lor}}

\newcommand{\Jc}{J_\text{c}}
\newcommand{\Jcstep}{J_\text{c}^\text{step}}
\newcommand{\Jcwell}{J_\text{c}^\text{well}}
\newcommand{\Jcpeak}{J_\text{c}^\text{Lor}}

\newcommand{\maxJcstep}{\text{max}[J_\text{c}^\text{step}/\Dconst J_\text{qb}]}
\newcommand{\maxJcpeak}{\text{max}[J_\text{c}^\text{Lor}/\Dconst J_\text{qb}]}

\newcommand{\maxCOPstep}{\text{max}[\text{COP}^\text{step}/\carnot]}
\newcommand{\maxCOPpeak}{\text{max}[\text{COP}^\text{Lor}/\carnot]}

\newcommand{\mumaxJcstep}{\left(\dmu/\gamma\right)|_{\text{max}J_\text{c}}}
\newcommand{\mumaxCOPstep}{\left(\dmu/\gamma\right)|_\text{maxCOP}}
\newcommand{\mumaxJcpeak}{\left(\dmu/\Gamma\right)|_{\text{max}J_\text{c}}}
\newcommand{\mumaxCOPpeak}{\left(\dmu/\Gamma\right)|_\text{maxCOP}}
\newcommand{\mumaxJcQSH}{\left(\dmu/\varepsilon_\perp\right)|_{\text{max}J_c}}
\newcommand{\Pabs}{P_\text{abs}}

\newcommand{\carnot}{\eta}
\newcommand{\COP}{\text{COP}}

\newcommand{\sigmaa}{\sigma_\alpha}

\newcommand{\kB}{k_\text{B}}
\newcommand{\vF}{v_\text{F}}

\newcommand{\Dconst}{D}

\newcommand{\ewell}{\varepsilon_\text{w}}
\newcommand{\epeak}{\varepsilon_\text{L}}
\newcommand{\VRB}{V}
\newcommand{\eperp}{\varepsilon_\perp}

\newcommand{\jsout}[1]{\expandafter\sout\expandafter{\textcolor{magenta}{#1}}}

\begin{document}

\title{Detailed study of nonlinear cooling with  two-terminal configurations of topological edge states}
%power and efficiency of a  quantum spin-Hall  refrigerator
%\textcolor{blue}{and other devices with similar transmission functions}
%\title{Boosting thermoelectric cooling in a  quantum spin-Hall  device}

	\author{Fatemeh Hajiloo}
	\thanks{These two authors contributed equally.}
	\affiliation{Department of Microtechnology and Nanoscience (MC2), Chalmers University of Technology, S-412 96 G\"oteborg, Sweden\looseness=-1}

\author{Pablo Terr\'en Alonso}
	\thanks{These two authors contributed equally.}
\affiliation{International Center for Advanced Studies, Escuela de Ciencia y Tecnolog\'ia and ICIFI,
Universidad Nacional de San Mart\'in, Avenida 25 de Mayo y Francia, 1650 Buenos Aires, Argentina}

	\author{Nastaran Dashti}
	\affiliation{Department of Microtechnology and Nanoscience (MC2), Chalmers University of Technology, S-412 96 G\"oteborg, Sweden\looseness=-1}

\author{Liliana Arrachea}
\affiliation{International Center for Advanced Studies, Escuela de Ciencia y Tecnolog\'ia and ICIFI,
Universidad Nacional de San Mart\'in, Avenida 25 de Mayo y Francia, 1650 Buenos Aires, Argentina}

	\author{Janine Splettstoesser}
	\affiliation{Department of Microtechnology and Nanoscience (MC2), Chalmers University of Technology, S-412 96 G\"oteborg, Sweden\looseness=-1}

\date{\today}

\begin{abstract}
We study the nonlinear thermoelectric cooling performance of a quantum spin Hall system. The setup consists of a nanomagnet contacting a Kramers' pair of helical edge states, resulting in a transmission probability with a rich structure containing peaks, well-type, and step-type
features. We present a detailed analysis of the impact of all these features on the cooling performance, based to a large extent on analytical results. We analyze the cooling power as well as the coefficient of performance of the device. 
Since the basic features we study may be present in the transmission function of other mesoscopic conductors, our conclusions provide useful insights to analyze the nonlinear thermoelectric behavior of a wide class
of quantum devices.
The combination of all these properties define the response of the quantum spin Hall setup, for which we provide some realistic estimates for the conditions limiting and optimizing its operation as a cooling device.  
\end{abstract}

%\pacs{ }
\maketitle

%%%%%%%%%%%%%%%%%%%%%%%%%%%%%%%%%%%%%%
\section{Introduction}
%%%%%%%%%%%%%%%%%%%%%%%%%%%%%%%%%%%%%%

Thermoelectric effects in the quantum regime is a very active avenue of research~\cite{Giazotto2006Mar,Benenti2017Jun,ozaeta2014,Marchegiani2020}.
While the device operation in this regime is mostly limited to ultra low temperatures and tiny output powers, it is intriguing to explore their prospects for  on-chip heat control, energy harvesting, and cooling in novel nanoscale systems. This would, for example, allow to avoid harmful heating of a delicate quantum device by local cooling in situ~\cite{Giazotto2006Mar}. A special asset is that their thermoelectric properties are coherently tunable to reach an optimal performance. 
It is hence of great interest to find appropriate quantum devices, which allow implementing such an on-chip cooling operation. Thermoelectric effects aiming at device cooling have recently been analyzed in quantum dots~\cite{Prance2009Apr,Zhang2015May,Entin-Wohlman2015Feb,Roura-Bas2018Nov,Venturelli2013Jun,Schulenborg2017Dec,Sanchez2018Dec} and quantum point contacts~\cite{vanHouten1992Mar,Whitney2013Aug,Whitney2015Mar,yamamoto17}, as well as in the edge states of quantum Hall and quantum spin Hall (QSH) systems~\cite{Rothe2012,Hwang2014Sep,Sanchez2015Apr,Vannucci2015Aug,Vannucci2016,Roura-Bas2018Feb,Mani2018Feb,boehling18,Sanchez2019Jun,Gresta2019Oct,Blasi2020}.
The  edge states in the latter two systems are topologically protected, being  helical Kramers' pairs with different spin polarizations  in the case of the QSH regime~\cite{kanemele2005,Bernevig1757,Koenig766,Roth294,laurens-2020}.

In a recent work, a QSH bar, in which the helical edge states are coupled through a nanomagnet, has been proposed as a versatile and tunable thermoelectric element~\cite{Gresta2019Oct}. The device was shown to contain the ingredients for
high-performance operation as a heat engine as well as for a high figure of merit. This is particularly interesting since 
this structures has been previously 
studied in relation to spintronics and topological quantum computing. In particular, in Refs.~\cite{Meng2014Nov,Arrachea2015Nov,Silvestrov2016May}, the interplay between  spin-torque and charge pumping was analyzed.
In combination to superconducting contacts, this structure has been investigated as a platform for  topological superconductivity~\cite{Fu2009Apr,Jiang2013Feb,Houzet2013Jul,Crepin2014May}.  
The additional use of such a device as thermoelectric refrigerator would hence be extremely beneficial once these principles are experimentally implemented. The previous analysis as a thermoelectric element~\cite{Gresta2019Oct} was however restricted to the linear regime of small voltage and temperature differences and did not analyse the cooling performance in detail.

 %%%%%%%%%%%%%%%%%%%%%%%%%%%%%%%%%%%%%%%%%%%%%
\begin{figure}[b]
	\centering
	\includegraphics[width=0.95\columnwidth]{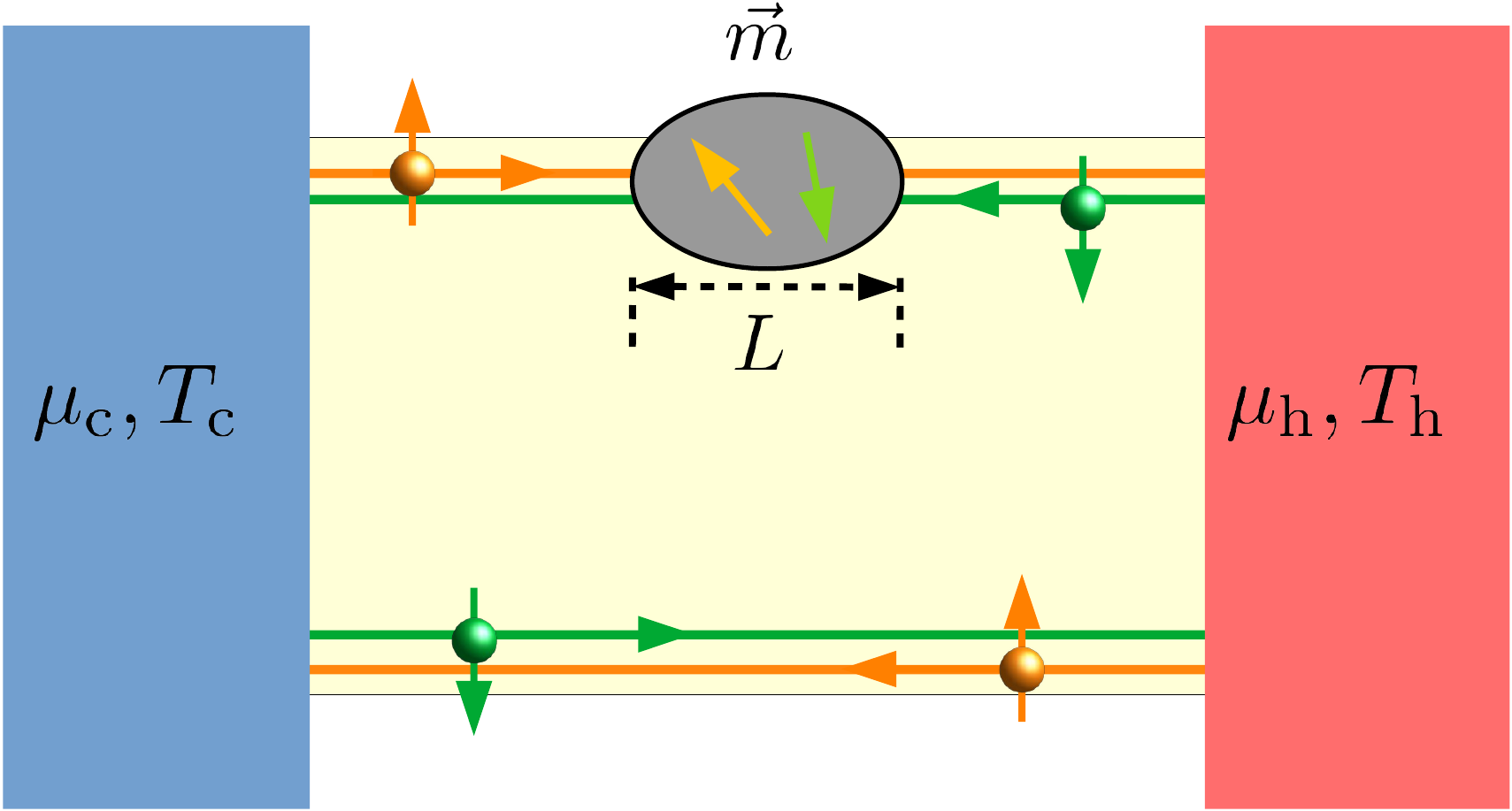}
	\caption{Sketch of the quantum spin Hall device in contact with cold and hot electronic reservoirs, with temperatures $\Tc,\Th$ and electrochemical potentials $\muc,\muh$. The reservoirs are connected by counter propagating helical edge states of opposite spin polarizations, indicated by orange and green; the horizontal arrows indicate the direction of propagation of the electrons. A magnetic island of length $L$ with magnetic moment $\vec{m}$ leads to backscattering accompanied by spin flips. }
	\label{fig:setup}
\end{figure}
 %%%%%%%%%%%%%%%%%%%%%%%%%%%%%%%%%%%%%%%%%%%%%

In the present paper, we study the full cooling performance of a two-terminal QSH device, reaching from the linear to the nonlinear regime of arbitrary temperature and voltage biases. The nonlinear regime allows for large cooling power and also includes the situation of cooling in the presence of large temperature differences between heat baths.  The specific setup, which we analyze, is sketched in Fig.~\ref{fig:setup} and introduced in detail in Sec.~\ref{sec:model}. It is a two-terminal setup, in which transport between the hot and cold electronic reservoir takes place via helical, spin-polarized topological edge states. Inter-edge backscattering  is induced at a metallic, magnetic island placed on top of one of the edges of the conductor.
Due to the induced spin-flips, the transmission probability of the conductor~\cite{Gresta2019Oct} has a complex and rapidly changing behavior as function of energy, see Fig.~\ref{fig:transmission}. This is a favorable situation for thermoelectric applications, which require particle-hole symmetry breaking, see e.g. Ref.~\cite{Benenti2017Jun} for a review. Interestingly, the transmission probability of our device has two different prominent characteristics, which arise from backscattering and interference effects. First, it shows an energy-gap with a (smoothed) step-like behavior to finite transmission; this is similar to the transmission probability of a quantum point contact (QPC). The performance of a QPC as thermoelectric refrigerator has been shown to have optimal efficiencies at fixed power~\cite{Whitney2015Mar,Whitney2013Aug}. The energy-intervals of finite transmission probability, furthermore, show oscillations as function of energy, which stem from the rigidness of the region inducing backscattering. The resulting peak-dip structure can be rather sharp depending on the length of the island. These features resemble the resonant features in the transmission probability of a quantum dot ~\cite{Prance2009Apr,Zhang2015May,Entin-Wohlman2015Feb,Roura-Bas2018Nov,Venturelli2013Jun,Schulenborg2017Dec,Sanchez2018Dec} or a Fabry-Perot interferometer~\cite{Sternativo2014Sep}. Sharp resonances have been shown to lead to thermoelectric efficiencies close to the Carnot limit in the linear-response regime~\cite{Hicks1993May,Hicks1993Jun,Mahan1996Jul,Heremans2013Jun}.

It can thus be expected that the complex transmission properties of the two-terminal QSH conductor of interest lead to similar cooling properties. 
We therefore start our analysis with an in-depth study of the cooling performance of conductors with transmission probabilities given by smoothed step functions and of conductors with transmission probabilities given by sharp resonances. 
This allows us to understand the complex cooling performance of the QSH device of interest in the full nonlinear regime of operation. Based on this analysis, we are able to identify the required parameter regimes to optimize the cooling power and efficiency of a QSH-based refrigerator.

However, this analysis is furthermore of importance for the cooling performance of a broad class of devices. Paradigmatic cases of transmission probabilities with some of these features are QPCs and quantum dots, as mentioned above. 
What is more, this strategy, which we present for the analysis of the cooling performance of the QSH device, can even be transferred to the analysis of other conductors with a similarly complex transmission probability. We explicitly show the example of a sharp, rectangular potential region in the Appendix~\ref{app:RB}. Also devices with edge-state based interferometers with side-coupled dots~\cite{Samuelsson2017Jun} or  molecular devices displaying interference effects leading to step-like transmission probabilities~\cite{Karlstrom2011Sep}, have been proposed to be exploited for thermoelectric effects. Furthermore, it is interesting to mention that in Corbino systems of the quantum Hall effect, diffusive transport through the Landau levels are accurately  described 
in terms of a transmission function presenting some of these key features \cite{corbino}.

The remainder of this paper is organized as follows. In Sec.~\ref{sec:setting}, we define the observables of interest for the cooling performance in terms of scattering theory, see Sec.~\ref{sec:cool-pow}, introduce the QSH device and all reference systems. see Sec.~\ref{sec:model}, and analyze the parameter range allowing for cooling, see Sec.~\ref{sec:range}. Finally we present results for the in-depth analysis of the cooling performance of conductors with a well-shaped, Sec.~\ref{sec:well}, step-shaped, Sec.~\ref{sec:step}, and resonant, Sec.~\ref{sec:peak}, transmission probabilities, and terminate with the QSH conductor at the focus of this work in Sec.~\ref{sec:QSH}. 
In Sec.~\ref{sec:conclusion}
we present a summary and conclusions.

%%%%%%%%%%%%%%%%%%%%%%%%%%%%%%%%%%%%%%%%%%%%%%%%%%%%%%%%%%%
\section{Observables for cooling performance and device properties}\label{sec:setting}
%%%%%%%%%%%%%%%%%%%%%%%%%%%%%%%%%%%%%%%%%%%%%%%%%%%%%%%%%%%

%%%%%%%%%%%%%%%%%%%%%%%%%%%%%%%%%%%%%%%%%%%%%%%%%%%%%%%%%%%
\subsection{Cooling power and coefficient of performance}\label{sec:cool-pow}
%%%%%%%%%%%%%%%%%%%%%%%%%%%%%%%%%%%%%%%%%%%%%%%%%%%%%%%%%%%

Without applying a potential bias $\dmu$ across a conductor, electronic heat currents always flow from the hot into the cold reservoir. In contrast,  we here analyze devices acting as refrigerators by using electric power to transfer heat from the cold to the hot reservoir. We focus on the coherent transport regime, where charge and heat currents can be fully  described in terms of a transmission function ${\cal T}(\varepsilon)$, see, e.g., Refs.~\cite{Blanter2000Sep,Moskalets2011Sep} for reviews and Ref.~\cite{Butcher1990Jun} for the treatment of energy currents.

The relevant quantities to be studied are, hence, the heat currents carried by electrons, $J_{\rm q, \alpha}$, flowing out of the cold and hot reservoirs, denoted respectively by $\alpha\equiv \text{c,h}$,   as well as the electrical power, $P$. We have
\begin{align}
\label{HeatCurrent}
J_{\rm q, \alpha}&=-\frac{1}{2\pi} \int_{-\infty}^{\infty} d\varepsilon \left(\varepsilon -\mu+\sigmaa\frac{\dmu}{2}\right) {\cal T}(\varepsilon) \sigmaa\df\\
	P&=\frac{\dmu}{2\pi} \int_{-\infty}^{\infty} d\varepsilon {\cal T}(\varepsilon) \df\label{eq:powerdef}\ .
\end{align}
We have introduced the abbreviation for the transport window, $\df=f_\text{c}(\varepsilon)-f_\text{h}(\varepsilon)$. 
The electronic occupation in the contacts is given by the Fermi distribution function 
\begin{equation}
	f_{\alpha}(\varepsilon)=\frac{1}{1+\exp\left[\left(\varepsilon -\mu+\sigma_{\alpha}\Delta\mu/2\right)/T_{\alpha}\right]},
\end{equation}
with $\sigma_{\text{c}}=-1$ and $\sigma_{\text{h}}=+1$. The potential difference $\dmu$ is defined with respect to an average potential $\mu$ as $\muc=\mu+\dmu/2$ and $\muh=\mu-\dmu/2$. Given the temperatures $T_\alpha$ entering the Fermi distributions, we define a temperature difference as $\dT=\Th-\Tc$. We set $\kB=\hbar=1$.

The energy-dependent transmission function of the device, ${\cal T}(\varepsilon)$, defines the transport properties of the conductor of interest. In the next section, we present these functions for a QSH system in contact with an extended magnetic region and for other prominent classes of conductors with similar properties.

We are interested in two quantities characterizing the device as a refrigerator. The first of these two quantities is the heat current flowing \textit{out of} the cold reservoir. We are interested in cooling, namely in the situation, when the heat current flowing out of the cold reservoir is positive. We then define  the \textit{cooling power} as
\begin{equation}
\Jc \equiv J_{\rm q, c}, \;\;\;\;\;\;\;\; \mbox{if}\;\; J_{\rm q, c} \geq 0.
\end{equation}
Otherwise, by hand, we set $\Jc\equiv 0$.
The upper bound for the cooling power that can be reached in this way has been found to be given by half of the Pendry quantum bound for heat currents~\cite{Pendry1983Jul,Whitney2013Aug,Whitney2015Mar}, $J_\text{qb}/2$, with 
\begin{equation}\label{eq:bound}
 J_\text{qb} = 
\frac{\pi^2 k_B^2}{6 h}  T_c^2.
\end{equation}

In order to characterize how efficient the refrigerator is, we need to compare the cooling power with the required absorbed electrical power, $P\equiv P_{\text{abs}}$. This defines the  \textit{coefficient of performance} (COP), which is given by 
\begin{equation}\label{eq:COP_def}
\text{COP}=\frac{\Jc}{\Pabs}.
\end{equation}
The coefficient of performance is bounded by the Carnot limit $\COP \leq \carnot$, with $\carnot= T_\text{c}/\left(T_\text{h}-T_\text{c}\right)$.  

For a complex device as the one studied here, see Fig.~\ref{fig:setup}, it is not obvious how to find the range of parameters that optimizes  $\Jc$ and/or $\COP$. In order to tackle this problem, we identify useful reference transmission functions, which the transmission probabilities of Fig.~\ref{fig:transmission}(a) can be compared to. This transmission probability as well as the ones of the reference conductors are introduced in detail in the following Sec.~\ref{sec:model}.

%%%%%%%%%%%%%%%%%%%%%%%%%%%%%%%%%%%%%%%%%%%%%%%%%%%%
\subsection{Quantum spin Hall and quantum Hall conductors with extended backscattering regions}\label{sec:model}
%%%%%%%%%%%%%%%%%%%%%%%%%%%%%%%%%%%%%%%%%%%%%%%%%%%%

We here present the transmission probabilities ${\cal T}(\varepsilon)$ that can be realized in  different devices hosting topological edge states. Transport in these devices is characterized by propagation along one-dimensional helical or chiral edge states. Backscattering in these devices is hindered by the topological properties, as long as no specific backscattering mechanism is introduced. Our main focus is on a quantum spin Hall device, in which backscattering at the same edge is enabled by spin flips due to the coupling to an extended magnetic region, see Fig.~\ref{fig:setup}. The resulting transmission probability, which can analogously be obtained due to inter-edge coupling in a constriction~\cite{Sternativo2014Sep}, has a complex energy-dependence. This, together with the possibility to tune to spectral regions with a broken electron-hole symmetry via a gate voltage, is a necessary condition for thermoelectric cooling and heat--work conversion in mesoscopic conductors.

A clearer understanding of the cooling performance of this device of interest can be obtained from a comparison with the cooling power of simpler transmission functions featuring steps and peaks as function of energy. In the following we introduce all of these transmission functions.

%%%%%%%%%%%%%%%%%%%%%%%%%%%%%%%%%%%%%%%%%%%%%%%%%%%%
\subsubsection{Quantum spin Hall conductor with magnetic island}\label{sec:model-qsh}
%%%%%%%%%%%%%%%%%%%%%%%%%%%%%%%%%%%%%%%%%%%%%%%%%%%%
We start with the device based on a quantum spin Hall conductor with a magnetic island, as sketched in Fig.~\ref{fig:setup}. Right- and left-moving electronic quasiparticles have spin orientations $\uparrow$  and $\downarrow$ and (Fermi) velocity $v_\text{F}$.
The grey region of length $L$ is assumed to have a uniform magnetic moment  $\vec{m}= \left(m_{ \perp} \cos \phi, m_{\perp} \sin \phi, m_{||}\right) $, with components $m_{||}$ (parallel)  and $m_{ \perp}$ (perpendicular) to the
natural quantization axis of the topological insulator. The traveling electrons feel the magnetic moment via an exchange interaction with strength $\mathcal{J}$. The effect of  a non-vanishing component $m_{ \perp}$ is to introduce backscattering at the edge states on the same side of the sample. This leads to the opening of a gap in the Dirac system of helical edge states. 

We employ a model for the helical edge states which neglects electron-electron Coulomb interactions. 
 Coulomb interactions are known to play a significant role in the cooling properties of quantum dots~\cite{Venturelli2013Jun,Schulenborg2017Dec,Sanchez2018Dec}. However, in the present context, it is important to notice that we focus on a single Kramers' pair of edge states in which case only the forward channel of the Coulomb interaction contributes. This component plays a relevant role in configurations where the edge states are 
 tunnel-coupled to other structures~\cite{Vannucci2015Aug, Vannucci2016, ronetti-2017,Roura-Bas2018Feb,Roura-Bas2018Nov}. On the other hand, the backward channels of the electron-electron interaction may also play a role in narrow constrictions connecting the  Kramers' pairs localized in opposite edges\cite{laurens-2020}. Nevertheless, for two-terminal configurations based on a single Kramers' pair like the one of Fig.~\ref{fig:setup} and also in the case of wide constrictions between two Kramers' pairs, the experimental evidence supports the description in terms of non-interacting edge states \cite{laurens-2020}.

%%%%%%%%%%%%%%%%%%%%%%%%%%%%%%%%%%%%%%%%%%%%%%%%%%%%
\begin{figure}%[h]
%\vspace{1.cm}
\begin{center}
\includegraphics[width=\columnwidth]{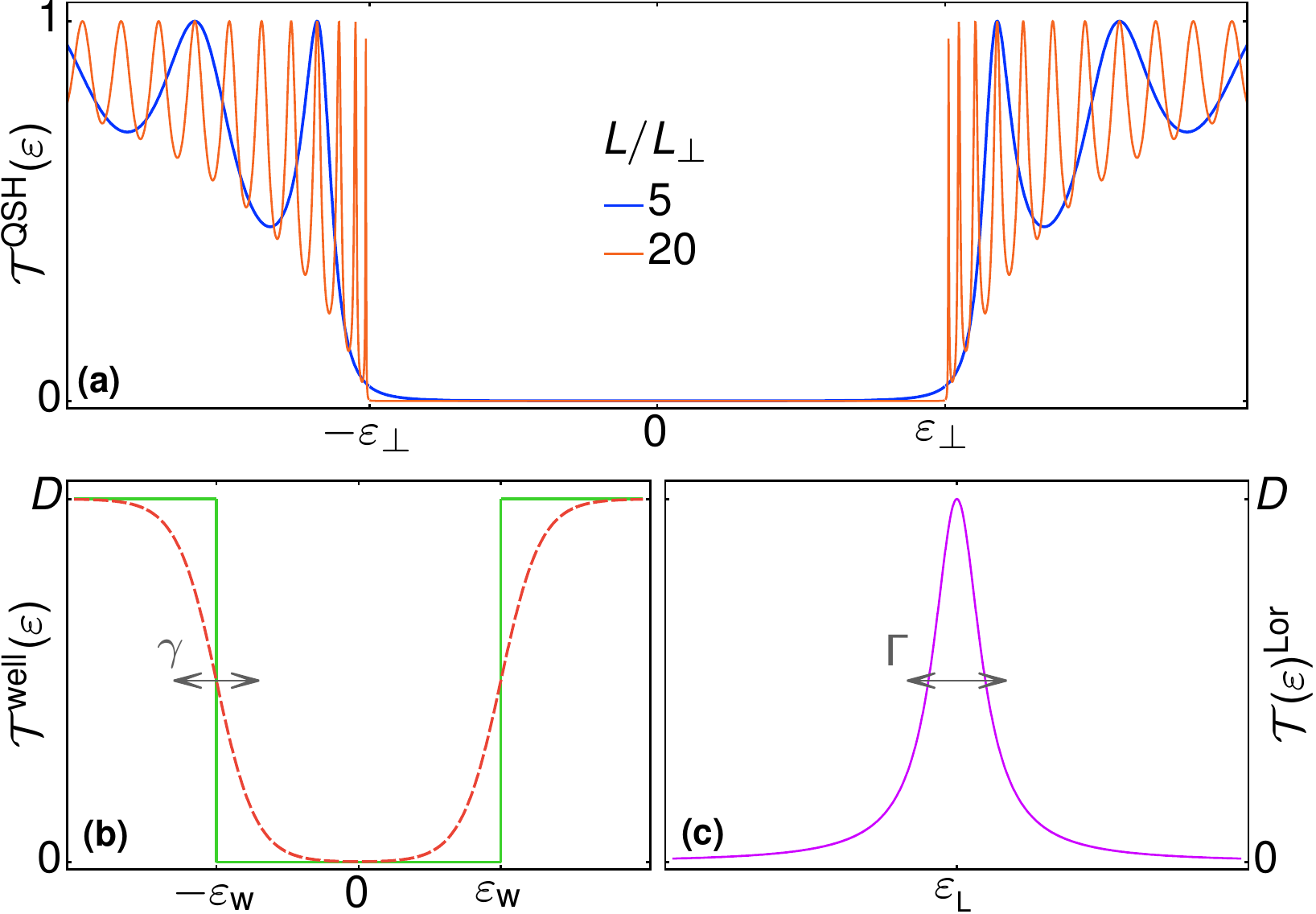}
\end{center}
\caption{Transmission probabilities for (a) the QSH device, as defined in Eq.~(\ref{eq:tau}). As reference, we show  in (b) a sharp and smooth, well-shaped transmission function, see Eq.~(\ref{eq:well}), as well as (c) a Lorentzian-shaped transmission probability, as defined in Eq.~(\ref{eq:peak}).}
\label{fig:transmission}
\end{figure}
%%%%%%%%%%%%%%%%%%%%%%%%%%%%%%%%%%%%%%%%%%%%%%%%%%%%

The resulting transmission probability of this system has been calculated previously~\cite{Gresta2019Oct} and is given by
\begin{align} 
\label{eq:tau}
\TQSH(\varepsilon) & = \frac{1}{1+\frac{\varepsilon_\perp^2}{|\varepsilon_\perp^2- \varepsilon^2|}|\sin(\lambda_\perp)|^2}.
\end{align}
 Here, we have introduced the complex, energy-dependent parameter $\lambda_\perp(\varepsilon)=  r_\perp(\epsilon)  L/ L_\perp$, with $L_\perp = \hbar \vF/\varepsilon_{\perp} $ and  
\begin{align}
    r_\perp(\epsilon)= & \left\{\begin{array}{ll}
   \sqrt{(\varepsilon/\varepsilon_{\perp})^2-1}\ ,      &  \text{if}\ \  |\varepsilon|\geq\varepsilon_\perp\\
 i\sqrt{1-(\varepsilon/\varepsilon_{\perp})^2}\ ,        & \text{if}\ \  |\varepsilon|<\varepsilon_\perp
    \end{array}\ \right. .
\end{align}
%\sqrt{\varepsilon^2- \varepsilon_{\perp}^2}/(\hbar v_F)$. 
Note that the transmission probability does not depend on the detailed orientation of the magnetic moment but only on the projection $m_{\perp}$ perpendicular to the direction of the natural quantization axis  of the material via the energy scale $\epsilon_\perp=\mathcal{J}m_\perp$. 

The transmission probability is shown in Fig.~\ref{fig:transmission}(a). It is
symmetric with respect to the Dirac point $\varepsilon = 0$. The magnetic moment introduces an effective coupling between the two Kramer's partners, leading to a gap of the order of $2\varepsilon_{\perp}$ in which the transmission probability is suppressed. This gap has smooth boundaries, depending on the inverse length of the magnetic island. 
For energies $|\varepsilon| > \varepsilon_{\perp}$, $\TQSH(\varepsilon)$ exhibits oscillations. Their amplitudes and positions also strongly depend on the inverse of the length of the magnetic island. More specifically, we find local maxima $\TQSH(\varepsilon_{2n+1}) =1$ and minima
  $\TQSH(\varepsilon_{2n}) = 1- \left(\varepsilon_{\perp}/\varepsilon_{2n}\right)^2 $
  at energies satisfying 
 \begin{align}
    \varepsilon_{\ell} & =\pm \varepsilon_{\perp}\sqrt{1+ \left( \frac{\pi(\ell+1) L_\perp}{2L}\right)^2}\ ,\label{eq:transmission_peaks}
 \end{align} 
 with $\ell=2n$ for minima and $\ell=2n+1$ for maxima with $n\in \mathbb{N}_0$. With this we define the spacing between the first two peaks,
 \begin{align}\label{eq:deltae}
     \Delta\varepsilon=\varepsilon_{3}-\varepsilon_{1},
 \end{align}
 which turns out to be an important energy scale for the cooling performance of the QSH device.
 
 Note that an analogous transmission probability is found for conductors, hosting topological edge states, with a sharp, extended constriction. In this case, weak coupling between counter-propagating edge states of the same spin orientation leads to backscattering and the opening of a gap. This has explicitly been shown for QSH devices with an etched constriction in Ref.~\cite{Sternativo2014Sep}. See, furthermore, Ref.~\cite{Bustos-Marun2013Aug} for an equivalent transmission in a topological system with driving.
 
The rich features of the transmission probability of this conductor make it an intriguing device for thermoelectric on-chip cooling. 
For the analysis of the cooling performance of this complex conductor, it is useful to compare to other conductors, which are characterized by similar, but simpler transmission probabilities. The transmission probabilities of interest are the following: (i) a well-shaped function of energy, which constitutes the envelope of the function given in Eq.~(\ref{eq:tau}). Furthermore, (ii) we will study a step-like function, representing each of the two sides of the well. And finally (iii) a peaked function of energy is of interest as it arises in the oscillations of $\TQSH$ at  $|\varepsilon|\geq\eperp$. 
Interestingly, several of the properties commented before also appear in a very simple system consisting in a rectangular potential barrier in  single-channel conductor. This is discussed
in Appendix \ref{app:RB}.

%%%%%%%%%%%%%%%%%%%%%%%%%%%%%%%%%%%%%%%%%%%%%%%%%%%%%%%%%%%%%
 \subsubsection{Envelope: well-shaped transmission probability}\label{sec:model-simpler}
%%%%%%%%%%%%%%%%%%%%%%%%%%%%%%%%%%%%%%%%%%%%%%%%%%%%%%%%%%%%%

The first transmission probability to compare with is a well-shaped function
\begin{equation}\label{eq:well}
 \Twell(\varepsilon) = \Dconst\left[1-\Theta_{\gamma}(\varepsilon + \ewell)+\Theta_{\gamma}(\varepsilon - \ewell)\right], 
\end{equation}
with
 \begin{equation}\label{eq:thetagamma}
\Theta_{\gamma}(\varepsilon)=\frac{1}{1+e^{-\varepsilon/\gamma}}.
 \end{equation}
The well has an extension $2\ewell$ and its steps have a smoothness quantified by the parameter $\gamma$, see Fig.~\ref{fig:transmission}(b). In the limit $\gamma \rightarrow 0$, the function in Eq.~(\ref{eq:thetagamma}) tends to the Heaviside-theta function $\Theta(\varepsilon)$. 
 We furthermore introduce a constant transmission prefactor $D\leq1$.
The transmission function of Eq.~(\ref{eq:well}) is the limiting case of Eq.~(\ref{eq:tau}) and Fig.~\ref{fig:transmission}(a) when we consider only the envelope and neglect the oscillations. As mentioned before, the origin of these oscillations is the sharpness of the region in which backscattering is induced.

% After Ref.~\onlinecite{buttiker}, it is usual to consider a phenomenological transmission function with the form of a smoothed step function for a constriction in QH edge states~\cite{Kheradsoud2019Aug}.  An aspect that has not been so far analyzed in this context is the fact that, not only the smoothed step function, but also its mirror for $\varepsilon<0$ plays a role in the non-linear transport behavior of the device. We will analyze this feature in detail below.

%%%%%%%%%%%%%%%%%%%%%%%%%%%%%%%%%%%%%%%%%%%%%%%%%%%%%%%%%%%%%
 \subsubsection{Step-like transmission probability}\label{sec:model-QPC}
%%%%%%%%%%%%%%%%%%%%%%%%%%%%%%%%%%%%%%%%%%%%%%%%%%%%%%%%%%%%%

The well-shaped transmission probability introduced above is simply composed of two steps as function of energy. Depending on the position of the transport window $\df$, only one of the steps might play a role for the characteristics of the cooling power of the QSH device. We therefore here introduce the function 
 \begin{equation}\label{eq:step}
 \Tstep(\varepsilon) = \Dconst\Theta_{\gamma}(\varepsilon - \ewell). 
\end{equation}

%%%%%%%%%%%%%%%%%%%%%%%%%%%%%%%%%%%%%%%%%%%%%%%%%%%%%%%%%%%%%
 \subsubsection{Peaked transmission probability}\label{sec:model-QD}
%%%%%%%%%%%%%%%%%%%%%%%%%%%%%%%%%%%%%%%%%%%%%%%%%%%%%%%%%%%%%
 
Finally, we introduce a peaked transmission probability, which can be used to model the sharp resonances under the well-shaped envelope of Eq.~(\ref{eq:tau}), 
\begin{equation}
    \Tpeak(\varepsilon)=D\frac{\Gamma^2}{\left(\varepsilon-\epeak\right)^2+\Gamma^2}.\label{eq:peak}
\end{equation}
We here decide to introduce a peak at energy $\epeak$ with a Lorentzian profile in energy with broadening $\Gamma$. The maximum of the peak in the transmission probability occurs at $\varepsilon=\epeak$ and is given by $\Dconst$.   The broadenings of these peaks in the transmission probability of the QSH conductor depend on the length of the magnetic island, and their positions are given by $\varepsilon_\ell$, with $\ell$ odd, as given by Eq.~(\ref{eq:transmission_peaks}). Most pronounced is the peak at $\varepsilon_1 \sim \varepsilon_\perp$, as can be seen in Fig.~\ref{fig:transmission}.

%%%%%%%%%%%%%%%%%%%%%%%%%%%%%%%%%%%%%%%%%%%%%%%%%%%%%%%%%%%%%
 \subsubsection{Model applicability}\label{sec:applicability}
%%%%%%%%%%%%%%%%%%%%%%%%%%%%%%%%%%%%%%%%%%%%%%%%%%%%%%%%%%%%%
 
 These model transmission functions, introduced in Secs.~\ref{sec:model-simpler} to \ref{sec:model-QD} are characteristic for other fundamental types of conductors as well. For example, we expect Eq.~(\ref{eq:well}) to be an appropriate description of devices, where the backscattering amplitude between counter-propagating edge states increases smoothly. This can occur in etched  constrictions~\cite{Sternativo2014Sep} with a smooth shape, as it is typical for quantum point contacts. Furthermore, the famous transmission probability of Eq.~(\ref{eq:step}) is the common description of the properties of a quantum point contact realized by a saddle point potential~\cite{Fertig1987Nov,Buttiker1990Apr}. Finally, the peak feature in the transmission probability of Eq.~(\ref{eq:peak}) is characteristic for example in quantum-dot devices, where it has been analyzed and exploited for cooling~\cite{Edwards1993Sep,Prance2009Apr,Whitney2015Mar,Sanchez2016Dec,Sanchez2019Jun}.  In particular, it is an appropriate description of an electronic Fabry-Perot interferometer, corresponding to a quantum dot realized by two subsequent quantum point contacts in quantum Hall devices, see e.g. Refs.~\cite{vanWees1989May,Chamon1997Jan,Ofek2010Mar,McClure2012Jun,Baer2015}.
 
Note, that in general care has to be taken when using these transmission probabilities for the calculation of \textit{nonlinear} transport properties. In the nonlinear transport regime, the energy-dependent transmission probability of a conductor is possibly influenced by the presence of large voltage differences via screening effects, see e.g. Ref.~\cite{Christen1996Sep,Meair2013Jan,Sanchez2016Dec,Texier2018Feb}. 
Importantly, in the case of the quantum spin Hall conductor with a magnetic island, which is the main motivation of the present work, the energy-dependence of the transmission probability stems from the magnetic properties of an extended metallic island. We expect this to lead to highly effective screening of charge accumulation in the vicinity of the region where backscattering is enabled by spin-flips induced by the magnetic moment. At distances larger than the short-range magnetic interaction length of this large magnetic domain, possible potential modifications due to large biases can not influence the transmission properties, since backscattering is inhibited in the absence of spin flips. For these reasons, we neglect possible small modifications of the potential landscape across the QSH scatterer in this paper.
 
For QPCs and quantum dots this approximation is in general not justified: in specific realizations or parameter regimes, screening effects can be neglected, vanish due to symmetry reasons~\cite{Kheradsoud2019Aug}, or additional gating can be used to restore the original transmission functions, see, e.g., Ref.~\cite{Benenti2017Jun}.
However, even when screening effects play a role, the analysis of the forthcoming sections is useful and relevant for the nonlinear thermoelectric behavior of these fundamental devices, because of two main reasons. On one hand, it provides a necessary starting point for the analysis of the more detailed description, where the screening potential is considered. On the other hand, the effect of screening is to introduce modifications in the energy dependence of the transmission probability: our study both gives an approximate result and guidance on cooling effects expected from induced modifications.

%\iffalse
%%%%%%%%%%%%%%%%%%%%%%%%%%%%%%%%%%%%%%%%%%%%%%%%%%%%%%%%%%%%%
%\begin{figure}[t]
%\begin{center}
%\includegraphics[width=\columnwidth]{figures/WELL_rigid_limits_mus.pdf}
%\includegraphics[width=\columnwidth]{figures/soft_well_limits.pdf}
%\caption{Range of parameters within which nonzero cooling power can be reached for a device characterized by the well-type transmission function  $\Twell(\varepsilon)$ defined in Eq.~(\ref{eq:well}) with $\gamma \rightarrow 0$. The shaded regions correspond to $\Jc\neq0$.
%The temperature of the coldest reservoirs is 
%(a) $\Tc = 0.05 \eperp$, (b)  $\Tc = 0.2 \eperp$, (c) $\Tc = 0.4 \eperp$, (d)  $\Tc = 0.6 \eperp$.
%\label{fig:hardwell_limits}}
%\end{center}
%\end{figure}
%%%%%%%%%%%%%%%%%%%%%%%%%%%%%%%%%%%%%%%

%%%%%%%%%%%%%%%%%%%%%%%%%%%%%%%%%%%%%%%
%\begin{figure}[t]
%\begin{center}
%\includegraphics[width=\columnwidth]{figures/boundary.pdf}
%\caption{ Same as Fig.~\ref{fig:hardwell_limits} for the chemical potential of the coldest reservoir fixed at $\muc=\eperp$ in a device with a well-type transmission function $\Twell(\varepsilon)$ defined in Eq.~(\ref{eq:well}) with $\gamma \rightarrow 0$ (a) and  (b). (c) corresponds to a 
% QSH device, described by $\TQSH(\varepsilon)$  with $L=20L_0$.
%\label{fig:softwell_limits}}
%\end{center}
%\end{figure}
%%%%%%%%%%%%%%%%%%%%%%%%%%%%%%%%%%%%%%%%%%%%%%%%%%%%%%%%%%%%%
%\fi

%%%%%%%%%%%%%%%%%%%%%%%%%%%%%%%%%%%%%%%%%%%%%%%%%%%%%%%%%%%%%
\subsection{Energy scales limiting the cooling range}\label{sec:range}
%%%%%%%%%%%%%%%%%%%%%%%%%%%%%%%%%%%%%%%%%%%%%%%%%%%%%%%%%%%%%
Before starting  the analysis of the cooling properties described by the different transmission functions, we recall that these functions are characterized by different energy scales. These are $\varepsilon_{\rm w}$ for the well-shaped case, 
 $\gamma$ for the smoothed step-type  function, and $\Gamma$ for the width of the Lorenzian peak. In our study we find it useful and natural to express the temperatures and chemical potentials in units of these scales. In the case of the
 QSH device, it is clear that $\varepsilon_{\perp}$ plays the role of $\varepsilon_{\rm w}$, while we define equivalent scales to $\gamma$ and $\Gamma$ associated to the smoothness of the gap closing and the width of the peaks, respectively. The non-linear behavior will manifest itself when the transport window, centered at $\mu$ and determined by $V$ and $\Delta T$, is  comparable with these scales or with $T_{\rm c}$.

We analyze here the parameter range in which cooling is actually possible in the QSH device.
We summarize the most prominent features of the  transmission probability Eq.~(\ref{eq:tau}). These are 
the well-type structure with a gap of size $2 \varepsilon_\perp$, which is at the heart of the Dirac nature of the topological edge states with
an induced backscattering, and  the structure of peaks and oscillations. This structure depends on the length of the magnetic island, which also determines the smoothness of the 
well-type function defined by the envelope connecting the maximums of $\TQSH$. 

\begin{figure}[t]
\begin{center}
\includegraphics[width=\columnwidth]{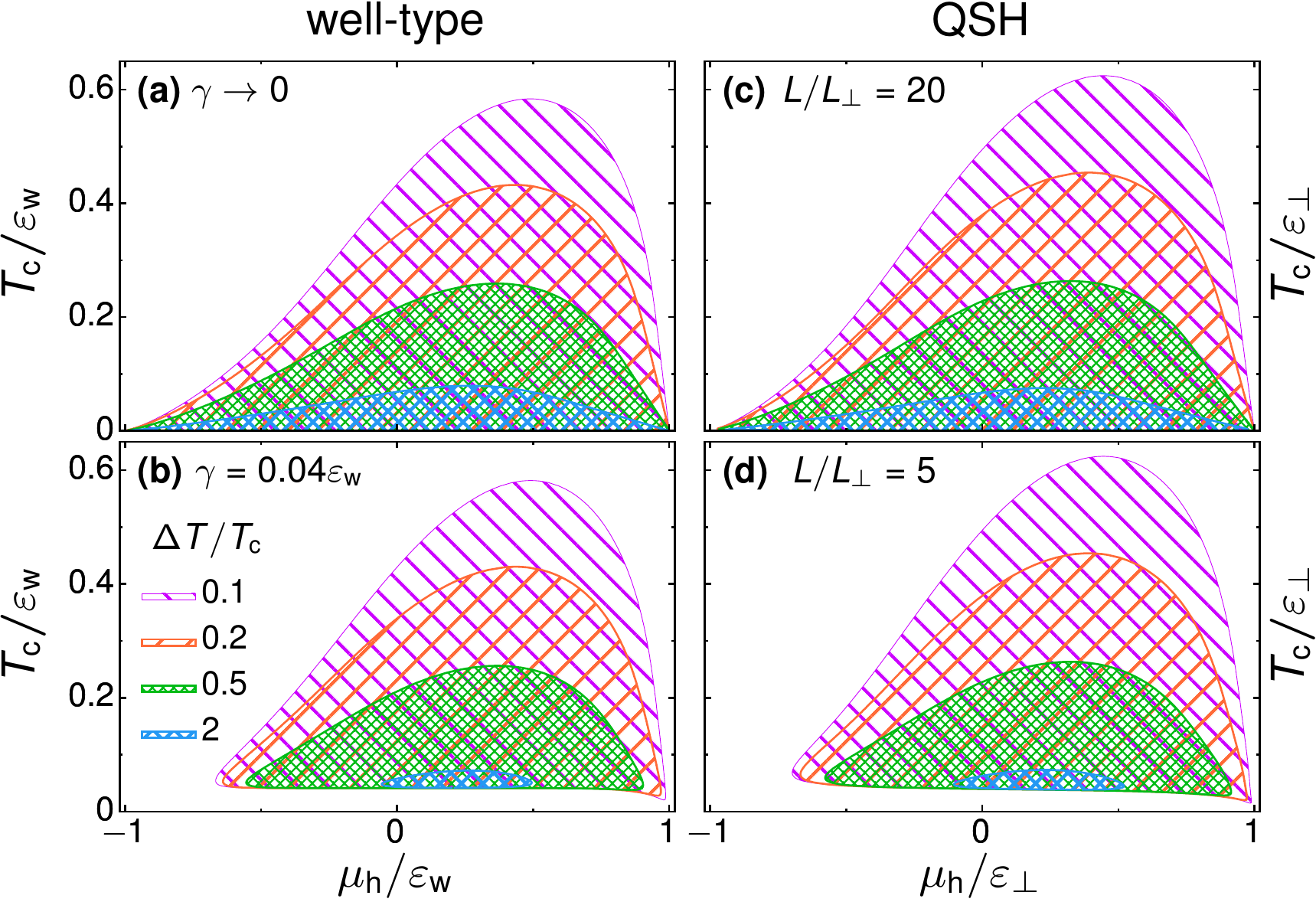}
\caption{Points in the $\{\Tc, \muh\}$ parameter space, where cooling is possible, namely $\Jc>0$. (a,b) device with a well-type transmission function $\Twell(\varepsilon)$ with $\gamma \rightarrow 0$ (a) and $\gamma=0.04\ewell$ (b), with fixed $\muc=\ewell$.
(c,d) QSH device at fixed $\muc=\varepsilon_\perp$, described by (c) $\TQSH(\varepsilon)$  with $L/L_\perp=20$ and (d) with $L/L_\perp=5$.
\label{fig:softwell_limits}}
\end{center}
\end{figure}

In Fig.~\ref{fig:softwell_limits} we present results for the range of $T_{\rm c}$ and $\mu_{\rm h}$ within which cooling is possible considering fixed ratios of $\Delta T/T_{\rm c}$ and $\mu_{\rm c}=\varepsilon_{\perp}$ for two different lengths $L/L_{\perp}$ of the magnet. We compare these results with those calculated with a well-type transmission probability $\Twell$ having a gap $2 \varepsilon_{\rm w}$, 
and   the same smoothness ratio as the QSH system, $\gamma/\varepsilon_{\rm w}=\gamma/\varepsilon_{\perp}$.
In the case of the QSH device, we define $\gamma$ from the slope of $\TQSH$ after the gap closing. This parameter is also related to 
the width $\Gamma$ of the first peak after the closure of the gap,
$\gamma= 2 \Gamma$. 
For  the shortest nanomagnet shown in Fig.~\ref{fig:softwell_limits}(d),  $L/L_{\perp}=5$, 
this corresponds to $\gamma/\varepsilon_{\perp}=0.04$
while $\gamma/\varepsilon_{\perp}\simeq 2 \times 10^{-3}$ in the case of $L/L_{\perp}=20$. 

We see that the results for the QSH device are strikingly similar to those obtained with a well-type function. In all the cases, we see that
the range for cooling shrinks for increasing  $\Delta T/T_{\rm c}$ and $\Delta \mu/\varepsilon_\perp$. This is a consequence of the fact that  when the transport window becomes large 
and covers a spectral range beyond the energy gap, the effects of particle-hole asymmetry in transport become weak, which is detrimental for thermoelectric cooling.

%%%%%%%%%%%%%%%%%%%%%%%%%%%%%%%%%%%%%%%
%%%%%%%%%%%%%%%%%%%%%%%%%%%%%%%%%%%%%%%%%%%%%%%%%%%%%%%%%%%%%

The boundaries for cooling, of course, depend on
the chemical potential of the cold reservoir. The ones shown in 
Fig. \ref{fig:softwell_limits} correspond to the specific value $\muc =\varepsilon_{\perp}$ for the QSH system  ($\muc=\ewell$ for the well-type case),
which is convenient for cooling because the energy-dependence of the transmission functions is most prominent in the vicinity of this point, as we will  discuss in detail in the next section.

%%%%%%%%%%%%%%%%%%%%%%%%%%%%%%%%%%%%%%%%%%%%%%%%%%%%%%%
\section{Cooling performance}\label{sec:results}
%%%%%%%%%%%%%%%%%%%%%%%%%%%%%%%%%%%%%%%%%%%%%%%%%%%%%%%

 In this section, we analyze in detail the full cooling performance of all three reference systems and exploit this to develop a complete understanding of the cooling performance of the QSH device.

%%%%%%%%%%%%%%%%%%%%%%%%%%%%%%%%%%%%%%%%%%%%%%%%%%%%%%%
\subsection{Conductor with well-shaped transmission}\label{sec:well}
%%%%%%%%%%%%%%%%%%%%%%%%%%%%%%%%%%%%%%%%%%%%%%%%%%%%%%%

As noticed before, a transmission function with the form of a sharp well with a gap of width $2\ewell$ is a good starting point to analyze the
cooling power. 
In the present section, we therefore study the cooling performance of a device described by the transmission function of Eq.~(\ref{eq:well}), first focusing on the limit of $\gamma\rightarrow0$ and then also on $\gamma\neq0$.

%%%% Intuitive discussion of the cooling physics

\begin{figure}[b]
    \centering
    \includegraphics[width=\columnwidth]{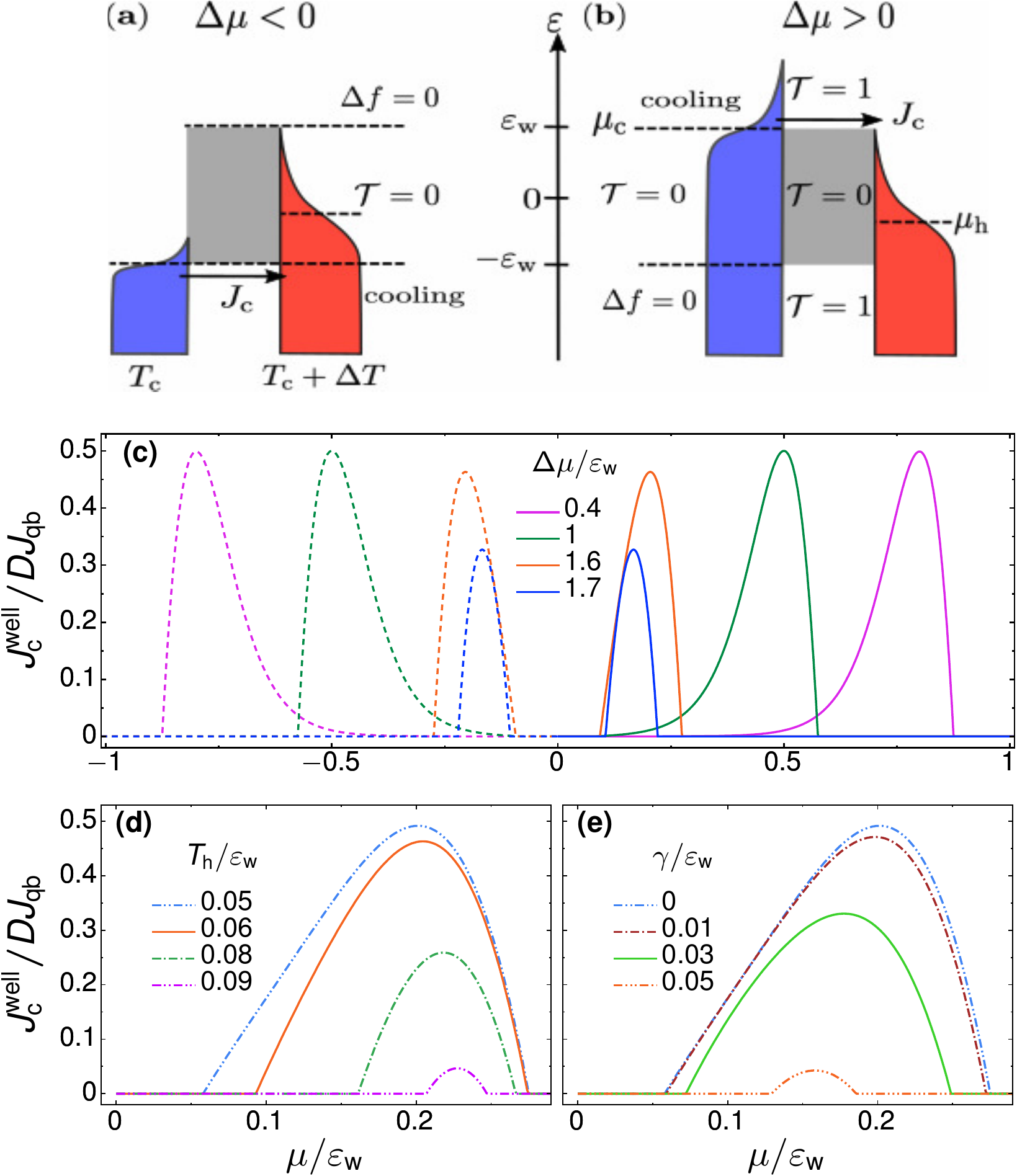}
    \caption{(a)-(b)~Energy landscape of a conductor with a well-shaped transmission probability connecting a hot and cold electronic distribution. The choice of potentials allows (a)~the hole-like excitations and (b)~the electronic excitations from the cold reservoir to leave to the hot in a specific energy window in a uni-directional manner. In the other energy intervals (separated by dashed lines) transport is either blocked due to $\Delta f=0$ or due to ${\cal T}=0$. (c)-(e)~Cooling power for a conductor with well-shaped transmission probability, Eq.~(\ref{eq:well}), as function of $\mu$ and parameters indicated in the panel legends. In panel~(c), dashed curves correspond to the negative values of parameters indicated by the legends of solid curves with the same colors. The otherwise fixed parameters are (c) $T_\text{c}/\ewell=0.05$ and $T_\text{h}/\ewell=0.06$, (d)~$T_{\text{c}}/\ewell=0.05$ and $\Delta\mu/\ewell=1.6$ and (e)~$T_\text{c}/\ewell=T_\text{h}/\ewell=0.05$ and $\Delta\mu/\ewell=1.6$.}
    \label{fig:CoolingPowerWell}
\end{figure}

The steps at $\varepsilon=\pm\ewell$ can both be exploited in order to cool down the distribution of an electronic contact as sketched in Fig.~\ref{fig:CoolingPowerWell}(a) and (b).
Let us first look at panel (b): here, the electrochemical potentials are placed with respect to the gap in the transmission function such that the electronic excitations in the cold contact (above $\muc$) can flow to the hot contact. In contrast the hole-like excitations (below $\muc$) can not be transferred due to the gap in the transmission.
This shows that the electrochemical potential of the colder distribution needs to be brought into the vicinity of one of the steps in order to get a large cooling power. The details of the cooling mechanism at such an individual steps are  discussed in Sec.~\ref{sec:sharp}.

At the same time, the electronic excitations of the hot reservoir are blocked from flowing into the cold reservoir by the gap as well. Importantly, below the gap, the occupations (Fermi functions) on both sides equal one, so that no hole transport (below $\muc$ and $\muh$) can take place. This shows that in order to achieve a good cooling performance, transport of only one particle species (electrons or holes) should take place at only one of the steps. In other words, electron-hole symmetry needs to be maximally broken.

The least symmetric situation is hence found, if only one of the two steps at $\pm\varepsilon_\text{w}$ lies in the transport window given by the difference in Fermi functions, $\df$, in Eqs.~(\ref{HeatCurrent}) and (\ref{eq:powerdef}).
For the temperatures and electrochemical potentials of the contacts, this requirement can be translated into the 
\begin{align}
    \text{\underline{requirement}}\ 1 &:  \Th  \ll 2\ewell-\dmu\label{eq:width_condition}
\end{align}
This condition means on one hand that $\Delta\mu$ should be smaller than the gap, but on the other hand that the smearing of the distribution functions should be small enough not to overlap with the region of high transmission below the gap.

The analogous situation for cooling by evacuating the hole-like excitations from the cold reservoir is shown in panel (a).
In the following, we consolidate the intuitive reasoning with a quantitative analysis.

%%%%%%%%%%%%%%%%%%%%%%%%%%%%%%%%%%%%%%%%
%%%%%%%%%%%%%%%%%%%%%%%%%%%%%%%%%%%%%%%%

%%%% Analytical results
We find the analytical expression for the cooling power for a conductor with a sharp, well-shaped transmission probability to be given by 
\begin{subequations}
\label{eq:sharp_well}
\begin{align}\label{Sharp_Well_J_1}
\Jcwell & = \Dconst\left[J_0+\Delta G(\ewell)-\Delta G(-\ewell)\right]\ .
\end{align}
Here, $J_0$ describes the standard contribution for Joule heating and heat conduction in the absence of a thermoelectric effect
\begin{align}\label{Sharp_Well_J_2}
J_0 & = -\frac{1}{2\pi}\left[ \frac{1}{2}\left(\dmu\right)^2 +\frac{\pi^2}{6}\left(\Th^2-\Tc^2\right)\right].
\end{align}
This term is always negative and hence decreases the total cooling power! Furthermore, we define $\Delta G(\varepsilon)=G_\text{c}(\varepsilon)-G_\text{h}(\varepsilon)$ where the function $G_{\alpha}$ is given as
\begin{align}
2\pi G_{\alpha}(\varepsilon)=&\left(\varepsilon-\muc\right) T_{\alpha}\text{ln}\left[1+e^{\left(\varepsilon-\mu+\sigma_{\alpha}\Delta\mu/2\right)/T_{\alpha}}\right]\nonumber \\
&+ T_\alpha^2\text{Li}_2\left[-e^{\left(\varepsilon-\mu+\sigma_{\alpha}\Delta\mu/2\right)/T_{\alpha}}\right],
\end{align}
\end{subequations}
with the dilogarithmic function $\text{Li}_2$. Electron-hole symmetry is maximally broken, if exactly one of the two functions  $\Delta G(\ewell)$ or $\Delta G(-\ewell)$ contributes to the cooling power, while the other does not. For example, in order to suppress $\Delta G(-\ewell)\approx0$ while $\Delta G(\ewell)\neq0$, we need to impose $\mu_\text{h/c}+\ewell\gg T_\text{h/c}$ together with $\ewell\approx\mu_\text{c}$. This is in agreement with the condition given in Eq.~(\ref{eq:width_condition}).

%%%%% Analysis based on plots
We finally, analyze the plots of the cooling power, obtained from Eqs.~(\ref{eq:sharp_well}), which are shown in Fig.~\ref{fig:CoolingPowerWell}(c)-(e). The cooling power is normalized with respect to its quantum bound. 
In panels~(c) and (d) of Fig.~\ref{fig:CoolingPowerWell}, we show the cooling power as a function of $\mu/\ewell$, while varying the parameters, $\Delta\mu$ and $T_\text{h}$, entering the condition in Eq.~(\ref{eq:width_condition}). First, in panel~(c), while fixing $T_\text{c}=0.05\ewell$ and $T_\text{h}=0.06\ewell$, we see that up to values of $\Delta\mu=1.6\ewell$ (namely $2\ewell-\Delta\mu=0.4\ewell$), the cooling power reaches its quantum bound. For even larger $\Delta \mu$, the cooling power gets suppressed due to transport of hole-excitations below $\mu$ from the hot to the cold reservoir. The same effect can be observed for negative potential biases: as soon as $\Delta\mu$ gets smaller than $-1.6\ewell$, the cooling power gets suppressed due to transport of electronic excitations above $\mu$ from the hot reservoir into the cold one. Panel~(d) then shows how an increase of $T_\text{h}$ reduces the value of the cooling power that can be reached, as well as the $\mu$-interval in which cooling is possible.

Up to here, we have focused on a \textit{sharp} well-shaped transmission probability. However, the smoothness $\gamma$ of steps also plays an important role.
For a smooth, well-shaped transmission probability, the condition given in Eq.~(\ref{eq:width_condition}) gets modified. The reason for this is that the smooth steps effectively decrease the size of the gap region, leading to a stronger constraint
\begin{align}
  \text{\underline{requirement}}\ 1' & :  \gamma,T_\text{h}\ll2\ewell-\Delta\mu.\label{eq:smooth_condition}
\end{align}
This additional constraint is relevant, only if the smoothness parameter $\gamma$ is at least of the same order as the temperature of the hot reservoir. This is also demonstrated in panel~(e) of Fig.~\ref{fig:CoolingPowerWell}. Note however, that the finite smoothness $\gamma$ also plays a major role for values of $\mu$ in the vicinity of $\ewell$. This will be further discussed in the following section. 

%%%%%%%%%%%%%%%%%%%%%%%%%%%%%%%%%%%%%%%%%%%%%%%%%%%%%%%
\subsection{Conductor with step-shaped transmission}\label{sec:step}
%%%%%%%%%%%%%%%%%%%%%%%%%%%%%%%%%%%%%%%%%%%%%%%%%%%%%%%

Under the conditions for sizable cooling power, set up in Sec.~\ref{sec:well} and given in Eqs.~(\ref{eq:width_condition}) and (\ref{eq:smooth_condition}), the conductor with a well-shaped transmission probability effectively behaves like a conductor with a single step in the transmission probability. For simplicity, we here focus on a step at $+\ewell$, where we can hence write the transmission probability as in Eq.~(\ref{eq:step}).
Equivalent considerations can be done for the step at $-\ewell$, where cooling takes place via hole-like transport. In the following, we analyze in detail the cooling performance of conductors starting with the sharp step, $\gamma\rightarrow0$ and proceeding to the smooth step, $\gamma\neq0$.

%%%%%%%%%%%%%%%%%%%%%%%%%%%%%%%%%%%%%%%%%%%%%%%%%%%%%%%
\subsubsection{Sharp step, $\gamma\rightarrow0$}\label{sec:sharp}
%%%%%%%%%%%%%%%%%%%%%%%%%%%%%%%%%%%%%%%%%%%%%%%%%%%%%%%

The mechanism leading to cooling for a sharp-step transmission probability has been briefly addressed in Sec.~\ref{sec:well} based on Fig.~\ref{fig:CoolingPowerWell}(a) and (b).
We start by analyzing the conditions that maximize the \textit{cooling power} of the conductor with a sharp step in the transmission probability.
If the electrochemical potential, $\mu_\text{c}$ is tuned to the edge of the transmission function, only particles occupying states above $\mu_\text{c}$ can leave the cold contact. This leads to the
\begin{align}
     \text{\underline{requirement}}\ 2 & :  \mu_\text{c}\approx \ewell \label{eq:position_condition}
\end{align}
At the same time, $\mu_\text{h}$ needs to be sufficiently below this edge, such that the occupation number of the hot contact is close to zero above the step. This means that the transport window defined by $\df$ needs to be large enough, such that the additional 
\begin{align}
     \text{\underline{requirement}}\ 3 & :T_\text{h}  \ll \Delta\mu \label{eq:step_cond}
\end{align}
is fulfilled.
Then, no back flow of excited particles from the hot contact can take place, which would otherwise lead to detrimental heating of the cold contact.

The analytic expression for the cooling power is directly found from Eq.~(\ref{eq:sharp_well}) as
\begin{align}\label{Sharp_step_J}
\Jcstep & = \Dconst\left[J_0+\Delta G(\ewell)\right]\ .
\end{align}
Indeed, Eqs.~(\ref{eq:position_condition}) and (\ref{eq:step_cond}) can be shown to maximize the contribution from $G_\text{c}(\ewell)$ and to minimize at the same time the contribution from $G_\text{h}(\ewell)$.

%%%%%%%%%%%%%%%%%%%%%%%%%%%%%%%%%%%%%%%%
  \begin{figure}[t]
    \centering
    \includegraphics[width=\columnwidth]{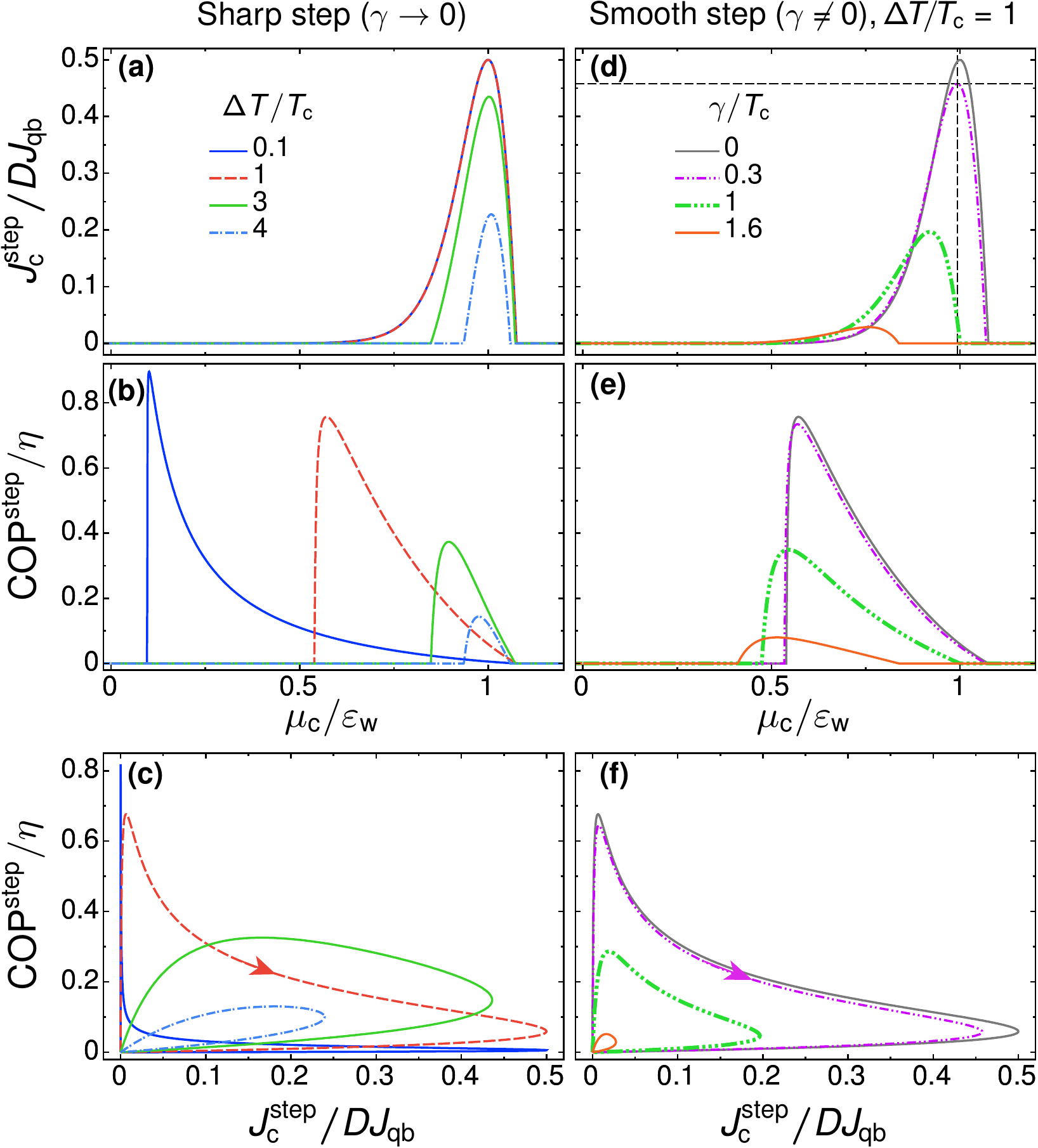}
    \caption{Cooling power and coefficient of performance for a conductor with sharp (left column with $\gamma\rightarrow0$) and smooth (right column with $\gamma\neq0$) step transmission probabilities, see Eq.~(\ref{eq:step}). We show the cooling power in (a) for different values of the temperature bias $\Delta T/T_\text{c}$ and in (d) for different values of smoothness $\gamma/T_\text{c}$ as indicated in the panel legends and the COP in panels (b) and (e). These four panels show results as function of $\mu_\text{c}$ with otherwise fixed parameters $\mu_\text{h}/\ewell=0$ and $T_{\text{c}}/\ewell=0.05$. The dashed black lines indicate the values $\Jcstep=0.457 D J_\text{qb}$ and $\muc=0.99\ewell$ as calculated from Eqs.~(\ref{eq:max_smooth}) and (\ref{eq:shift_smooth}). In panels (c) and (f) cooling power and COP are shown while $\Delta\mu$ is varied at fixed $\mu/\ewell=0.5$ and $T_{\text{c}}/\ewell=0.05$. 
    %\textcolor{red}{In panel (f) $\Delta T/T_\text{c}=1$}.
    \label{fig:Step}}
\end{figure}
%%%%%%%%%%%%%%%%%%%%%%%%%%%%%%%%%%%%%%%%

The two above discussed criteria for an optimization of the cooling power are furthermore demonstrated in Fig.~\ref{fig:Step}(a). Here, the electrochemical potential of the hot reservoir is chosen as energy reference, $\mu_\text{h}=0$. First of all, we see a peak in the cooling power at $\mu_\text{c}=\ewell$. As long as  $\Delta T/\ewell$ is small, the cooling power as a function of $\mu_\text{c}$ is exponentially suppressed with decreasing $f_\text{c}(\ewell)$ as $\mu_\text{c}/\ewell$ decreases. This behavior is independent of the distribution in the hot reservoir as long as $T_\text{c,h}\ll \varepsilon_\text{w}$. In contrast, if $\Delta T$ increases such that also $f_\text{h}(\ewell)$ becomes sizable, the region, in which the cooling power is finite, is reduced to a smaller interval of $\mu_\text{c}$ around $\ewell$. This is due to the flow of excited particles from the hot to the cold reservoir and the Joule heating going along with it. At the same time, the maximum value that the cooling power can take is reduced (as long as $\mu_\text{h}$ is fixed, as is the case here). This can be attributed to a decrease of the difference of occupation in the two contacts, $\df$ in the vicinity of the step in the transmission probability $\varepsilon\approx\ewell$. 
In summary, the cooling power can hence reach its bound as long as the condition $T_\text{h}\ll\Delta\mu$, see Eq.~(\ref{eq:step_cond}), is fulfilled.

When optimizing the performance of a cooling device, not only the cooling power plays a role, but one might equally be interested in the efficiency of the refrigerator. This is characterized by the COP, given in Eq.~(\ref{eq:COP_def}). The COP for the conductor with a sharp-step in the transmission probability is shown in  Fig.~\ref{fig:Step}(b).
The maximal value of the COP is typically found for values of the electrochemical potentials, for which the cooling power is finite, but small. The COP can however only be sizable at small cooling powers if at the same time also the absorbed power,
\begin{align}\label{Sharp_P}
P^{\text{step}}= \frac{\Dconst\Delta \mu^2}{2\pi}&\left[1 +\sum_{\alpha=\text{c,h}}\sigma_\alpha\frac{T_\alpha}{\Delta\mu}\text{ln}\left[f_{\alpha} (\ewell)\right]
\right]\ ,
\end{align}
is vanishingly small.
This is the case for a suppressed \textit{particle} flow between the two reservoirs. In the limit of small $\Delta T/T_\text{c}$, where particle flow from the hot to the cold reservoir is suppressed, the regions of zero cooling power and vanishing particle flow almost coincide, leading to the sharp peak in the COP. For increasing $\Delta T/T_\text{c}$, the position of maximum COP is furthermore influenced by Joule heating making the dependence of COP on $\mu_\text{c}$ smoother.

Due to the different behavior of the cooling power and the COP, it is not only of interest when one \textit{or} the other is maximal. We rather also want to find out under which conditions \textit{both} the cooling power and COP take sizable values. This is shown in the so-called lasso plots in Fig.~\ref{fig:Step}(c). These plots show the COP at different values of the cooling power, while the applied bias $\Delta\mu$ is changed in the direction indicated by the arrow. This clearly shows that ideal values for cooling power and COP are found for small temperature bias---however, they occur at different voltage bias, $\Delta \mu$. In other words, if $\Delta T/T_\text{c}\ll1$, the cooling power reaches its theoretical maximum $\Dconst J_\text{qb}/2$ at $\text{COP}\rightarrow 0$ and a COP close to the Carnot limit is reached at $J_\text{c}\rightarrow0$. In contrast, deep in the nonlinear response regime, $\Delta T,\Delta\mu>T_\text{c}$, COPs of approximately 30$\%$ of the Carnot value can be reached when the cooling power is about half of its maximum value $\Dconst J_\text{qb}/2$.

%%%%%%%%%%%%%%%%%%%%%%%%%%%%%%%%%%%%%%%%%%%%%%%%%%%%%%%
\subsubsection{Smooth step, $\gamma\neq0$}\label{sec:smooth}
%%%%%%%%%%%%%%%%%%%%%%%%%%%%%%%%%%%%%%%%%%%%%%%%%%%%%%%

If the transmission probability of the conductor gets the form of a smooth step, $\gamma\neq0$, the cooling performance is modified with respect to the case discussed above, see right column of Fig.~\ref{fig:Step}. The following observations can be made here. First of all, both the cooling power and the COP get reduced with increasing smoothness of the barrier as long as the electrochemical potential of the hot contact $\mu_\text{h}$ is fixed, as in the upper panels of Fig.~\ref{fig:Step}. Second, the position of the  maximum of the cooling power as function of the electrochemical potential of the cold reservoir is shifted away from $\mu_\text{c}=\ewell$ to smaller values, in contrast to the requirement for sizable cooling power for conductors with a sharp step in the transmission probability given in Eq.~(\ref{eq:position_condition})!
Intuitively, this can be understood in the following way. The slightly increased value of the cooling power at small $\mu_\text{c}$ (of the order of $\ewell/2$ in Fig.~\ref{fig:Step}(d)), stem from the energy dependence of the transmission probability, which for non-vanishing $\gamma$ extends into the gap region. The value of the cooling power for $\mu_\text{c}$ approaching $\ewell$ is instead suppressed, since the smooth transmission probability makes the transmission less ideal for the evacuation of electron-like excitations. At the same time, the barrier becomes opaque for hole-like excitation of the cold reservoir. If at those energies the occupation of the hot contact is finite, this in addition allows for a particle flow from the hot to the cold reservoir, thereby further reducing the cooling power. 
%{\color{violet} I'm not so sure about the last sentence. Of course with sufficiently large $\gamma$ the effects of the hot reservoir will appear. But both the extension of the cooling region and the changes in the peak (shift and value) are effects that occurs by just consider the step and the coldest reservoir's Fermi function. In fact I'm not sure that you're seeing any effect due to the hot reservoir at all in Fig.~\ref{fig:SmoothStep}(a) (if I understand correctly you're setting $\mu_h=0$ and $T_h=0.1\ewell$, which accordingly to Fig~\ref{fig:SharpStep} is almost invisible, even with $\gamma/T_c=1$).}

In the limit of small smoothness parameter $\gamma\ll T_\text{c}$, we find analytic expressions for these modifications.
For $\gamma \ll T_\text{c}$, the expression for the transmission function, Eq.~(\ref{eq:step}), which enters the cooling power in Eq.~(\ref{HeatCurrent}), can be expanded~\cite{Lukyanov1995Jan} to leading order in $\gamma$,
\begin{equation}
\Tstep(\varepsilon)
\rightarrow
\Dconst \left[ \theta(\varepsilon-\ewell)
+	
\gamma^{2}
\dfrac{\pi^2}{6} \dfrac{d }{d\varepsilon} \delta(\varepsilon-\ewell)\right].
\end{equation}
Using this expression in Eq.~(\ref{HeatCurrent}), the cooling power is then found to be 
\begin{align}\label{eq:smooth_step_J}
\Jcstep\rightarrow & \Jcstep(\gamma=0)-\gamma^2 \dfrac{\Dconst\pi}{12} \: \frac{d }{d\ewell} \left[ (\ewell-\mu_\text{c}) \:  \Delta f(\ewell)\right],
\end{align}
where $\Jcstep(\gamma=0)$ is given by Eq.~(\ref{Sharp_step_J}).  For example, for the parameter set used in  Fig.~\ref{fig:Step}(d), namely for $\mu_\text{h}=0,T_\text{c}/\ewell=0.05$ and $\Delta T/T_\text{c}=1$, this implies for the cooling power at $\mu_\text{c}=\ewell$,
\begin{align}\label{eq:max_smooth}
\left.\frac{\Jcstep}{\Dconst J_\text{qb}}\right|_{\mu_\text{c}=\varepsilon_\perp}\rightarrow\left.\frac{\Jcstep(\gamma=0)}{\Dconst J_\text{qb}}\right|_{\mu_\text{c}=\ewell}-\frac{1}{2}\left(\frac{ \gamma}{T_\text{c}} \right)^{2}.
\end{align}
From Eq.~(\ref{eq:smooth_step_J}), we can also derive the value of $\mu_\text{c}$ at which the  cooling power takes its maximum in the limit $\gamma/T_\text{c} \ll 1$. The result replaces requirement 2 in Eq.~(\ref{eq:position_condition}) for a small smoothness parameter $\gamma/T_\text{c}$
\begin{align}\label{eq:shift_smooth}
     \text{\underline{requirement}}\ 2' & : \mu_\text{c}\approx\ewell -\left(\dfrac{\gamma}{T_\text{c} }\right)^2 \:   \dfrac{1+4y }{1-2x} T_\text{c}\:,
\end{align}
where $x=f_\text{h}(\ewell)$ and $y=T_\text{c} \partial_{\ewell}f_\text{h}(\ewell)$. By choosing $\gamma=0.3T_\text{c}$, $T_\text{c}/\ewell=0.05 $ and $\Delta T/T_\text{c}=1$, the maximum cooling power, $\Jcstep=0.457 \Dconst J_\text{qb}$,   occurs at $\mu_\text{c}=0.99\ewell$. For a smoothness parameter $\gamma/T_\text{c}=0.3$ [purple-dashed curve in  Fig.~\ref{fig:Step}(d)], both the maximum cooling power, as well as the electrochemical potential $\mu_\text{c}$ at which it occurs, are indicated by thin red lines as they were obtained from Eqs.~(\ref{eq:max_smooth}) and (\ref{eq:shift_smooth}). 

The quadratic nature of the corrections to the cooling power for small $\gamma/T_\text{c}$ also show that in the limit where the temperatures of the contacts, $T_\text{c}$ and $T_\text{h}$, are much larger than the smoothness of the transmission function, $\gamma\ll T_\text{c},T_\text{h}$, the modifications with respect to the sharp barrier are small. In particular, in this limiting case, the conditions given in Eqs.~(\ref{eq:position_condition}) and (\ref{eq:step_cond}) continue to hold, representing the situation indicated in Fig.~\ref{fig:CoolingPowerWell}(a).

However, not only the cooling power, but also the COP is affected by an increasing smoothness of the transmission probability. More specifically, large values of the COP, which occur at suppressed cooling power are shifted to smaller potential biases with increasing smoothness parameter $\gamma$.
These properties also lead to the fact that the range in which the COP is close to its ideal value reduces to a window, where the cooling power is small with respect to the quantum bound. This can be seen in the lasso plots in Fig.~\ref{fig:Step}(f), where the line in the vicinity of large COPs is much more peaked compared to the one for a transmission probability with $\gamma\rightarrow 0$ in Fig.~\ref{fig:Step}(c). 

Before closing this subsection, we would like to draw the attention to the similar cooling properties of the  conductor with a step-type transmission probability, see Figs.~\ref{fig:Step}(a) and (d), and the rectangular barrier, see Fig.~\ref{appfig:RBmodel}(c) in Appendix \ref{app:RB}.

%%%%%%%%%%%%%%%%%%%%%%%%%%%%%%%%%%%%%%%%%%%%%%%%%%%%%%%
\subsubsection{Maximum cooling power and maximum COP for sharp and smooth step}\label{sec:max}
%%%%%%%%%%%%%%%%%%%%%%%%%%%%%%%%%%%%%%%%%%%%%%%%%%%%%%%

%%%%%%%%%%%%%%%%%%%%%%%%%%%%%%%%%%%%%%%%
  \begin{figure}[t]
    \centering
    \includegraphics[width=\columnwidth]{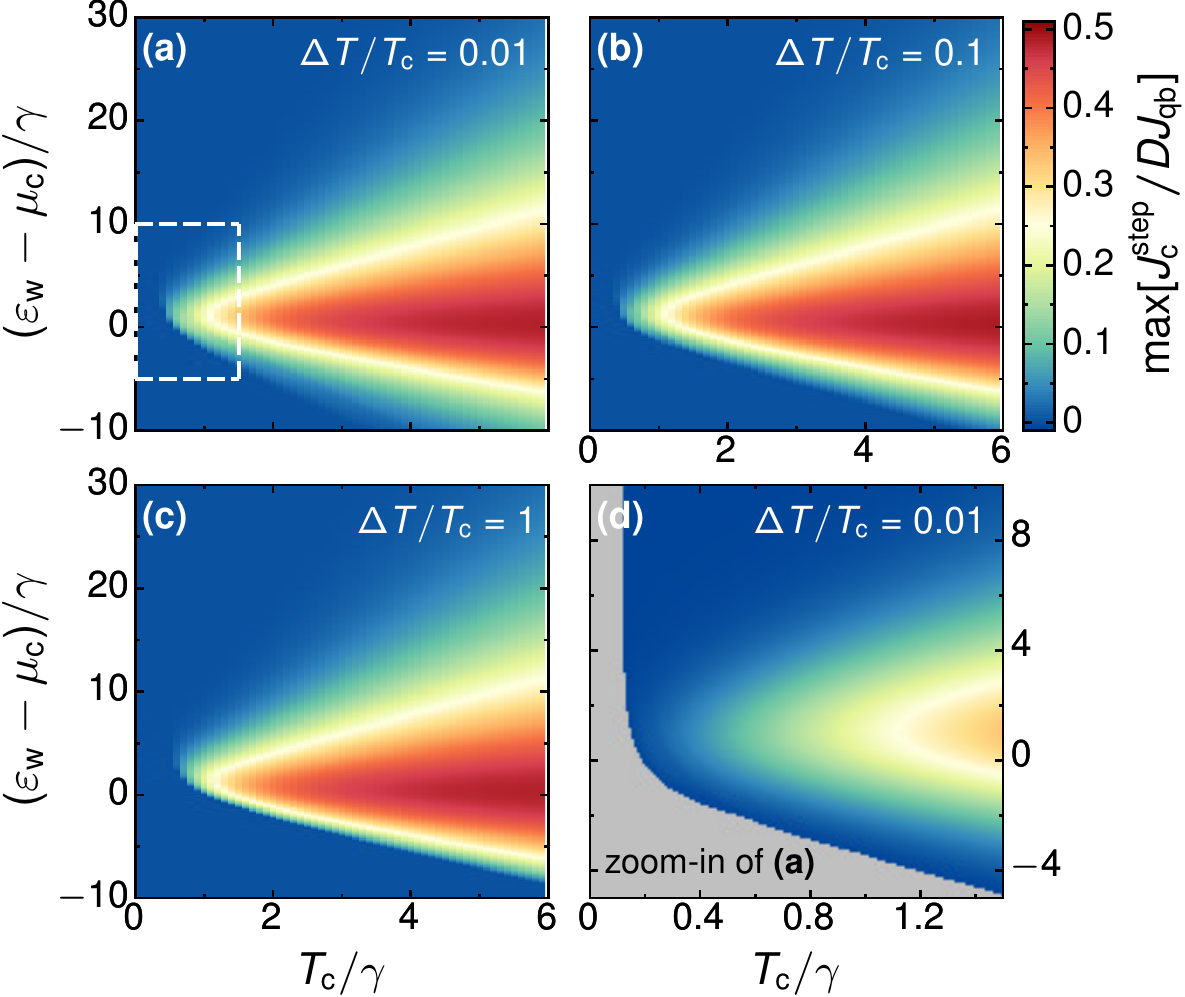}
    \caption{Normalized cooling power maximized over $\Delta \mu$, $\maxJcstep$, for a conductor with smooth-step transmission probabilities, Eq.~(\ref{eq:step}), as function of the step position, $\ewell$, with respect to the electrochemical potential of the cold reservoir, $\mu_\text{c}$, and the ratio between lowest temperature and step smoothness, $T_\text{c}/\gamma$. Panels~(a) to (d) differ by the temperature bias $\Delta T/T_\text{c}$; panel~(d) is a zoom-in of panel~(a) as indicated by the white-dashed rectangle.
    \label{fig:MaxJc_QPC}}
\end{figure}
%%%%%%%%%%%%%%%%%%%%%%%%%%%%%%%%%%%%%%%%

In the previous discussion of the cooling performance for a step-shaped transmission probability, we have seen different parameter sets, for which the cooling power reaches its quantum bound and where the COP reaches the Carnot limit. In the present subsection, we analyze the maximum values $\maxJcstep$ and $\maxCOPstep$, which these functions can take for any potential bias $\Delta \mu$.  

In Fig.~\ref{fig:MaxJc_QPC}, we show the maximum cooling power, $\maxJcstep$, as function of the position of the step with respect to the electrochemical potential of the cold reservoir, $\ewell-\mu_\text{c}$, and of the ratio between the temperature of the cold reservoir and the step smoothness, $T_\text{c}/\gamma$, for different temperature biases. 
These figures show that the cooling power is largest, and can even reach the quantum bound, in a region around $\mu_\text{c}\approx\ewell$, as expected from the discussion of Figs.~\ref{fig:CoolingPowerWell} and \ref{fig:Step}(a). Here, we find that it is the ratio between $T_\text{c}$ and $\gamma$ that determines the size of this region: The larger $T_\text{c}/\gamma$ gets, the larger cooling powers can be reached. In the low-temperature regime, the maximum cooling power is suppressed. How strong this suppression is, does however also depend on the temperature bias: In the linear-response regime, cooling---at very small cooling powers---is possible down to the zero-temperature limit, see panel~(a) and (d). The temperature bias is also found to limit the region of negative $\ewell-\mu_\text{c}$ in which cooling is possible. The reason for this is the following: if $\mu_\text{c}>\ewell$, hole-like excitations are created in the cold reservoir leading to heating. At energies, where $\mathcal{T}(\varepsilon)\neq 0$ this can only be hindered if at the same time $\Delta f(\varepsilon)$ is small. At large $T_\text{h}$, this however implies a heat flow of electron-like excitations from the hot reservoir into the cold, which hinders that cooling takes place in the cold reservoir at all.

%%%%%%%%%%%%%%%%%%%%%%%%%%%%%%%%%%%%%%%%
 \begin{figure}[t]
    \centering
    \includegraphics[width=1\columnwidth]{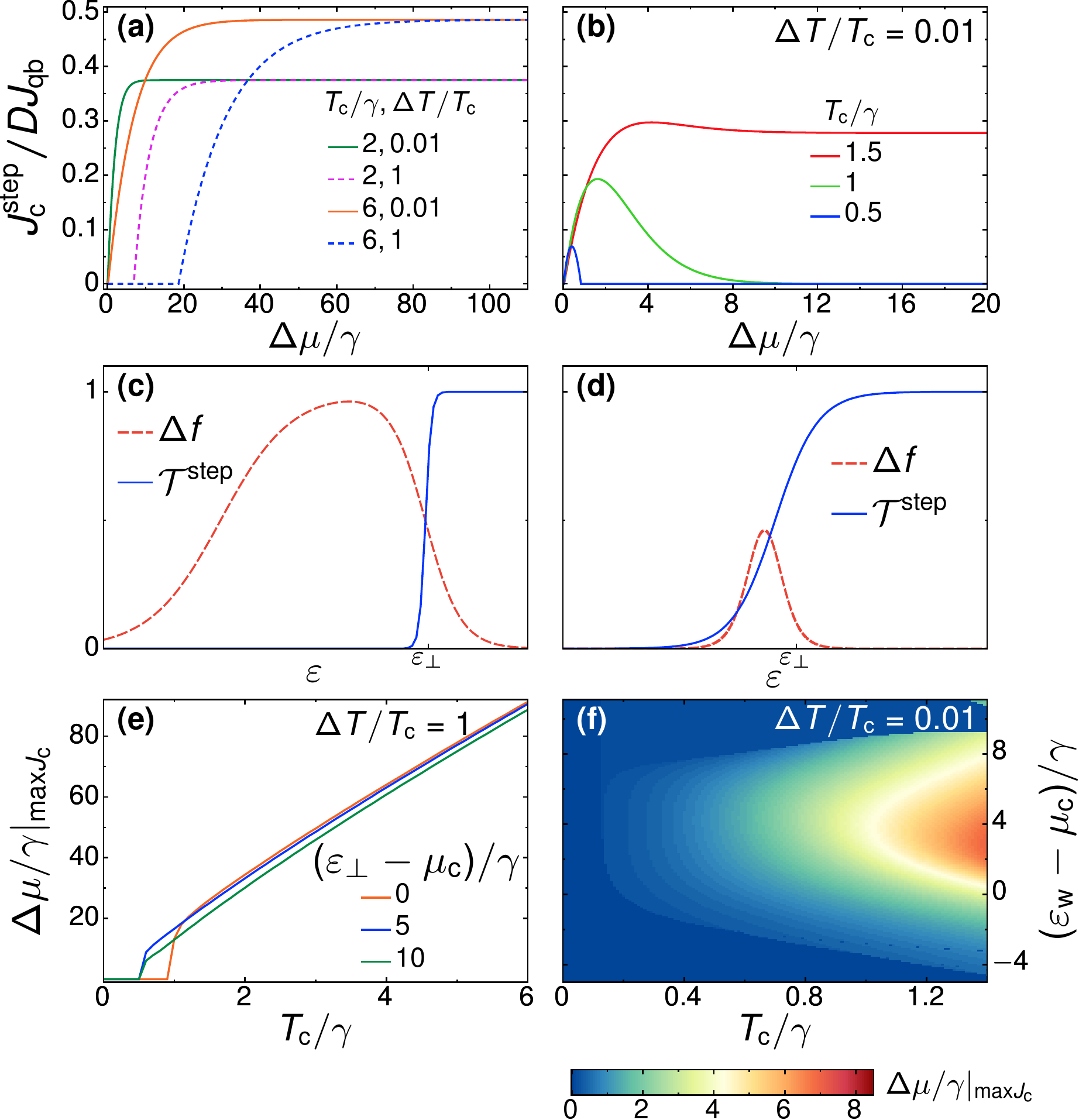}
    \caption{Analysis of the bias voltage $\mumaxJcstep$, at which the maximum cooling power, shown in Fig. (\ref{fig:MaxJc_QPC}), is reached. The left column shows the high-temperature regime, and the right column the low-temperature regime. Panel~(a) and (b) show the cooling power as a function of $\Delta \mu$ at fixed $\ewell=\muc$ and for different temperatures, as indicated in the legends. Panel~(c) and (d) show the transport window and the transmission probability that optimize the cooling power, for~(c) $T_{\text{c}}/\gamma=6$, $\Delta T/T_{\text{c}}=1$, $\Delta\mu/\gamma=70$ and (d) $T_{\text{c}}/\gamma=0.5$, $\Delta T/T_{\text{c}}=0.01$, $\Delta\mu/\gamma=1$. (e) Potential bias at which the maximum cooling power is reached when $\Delta T/T_\text{c}=1$ (obtained when the derivative of $J_\text{c}/D_0J_\text{qb}$ with respect o $\Delta\mu$ goes down to $10^{-4}/\gamma$). (f) Potential bias at maximum cooling power when $T/T_\text{c}=0,01$.
    \label{fig:DeltamuforMaxJc}}
\end{figure}
%%%%%%%%%%%%%%%%%%%%%%%%%%%%%%%%%%%%%%%

Fig.~\ref{fig:DeltamuforMaxJc} deals with the bias at which the cooling power is maximized. In particular we show here that the mechanism leading to maximum cooling power is very different in the low-temperature regime (requiring small potential biases leading to a linear-response behavior) and in the high-temperature regime (requiring large potential biases leading to nonlinear-response behavior). In the high-temperature regime, see the left column of Fig.~\ref{fig:DeltamuforMaxJc}, namely for $T_\text{c}, T_\text{h}\gg\gamma$, the maximum cooling power is reached, if only the electrochemical potential of the cold reservoir, $\mu_\text{c}$ is in the vicinity of $\ewell$, while electrons from the hot reservoir are fully blocked from transport thanks to a large potential bias, see panel~(c). The maximum value is reached asymptotically, as panel~(a) shows. 
In this case, the larger the potential bias, the larger the cooling power. Panel~(e) of Fig.~\ref{fig:DeltamuforMaxJc} demonstrates a steep linear increase of $\mumaxJcpeak$ as function of $T_\text{c}$.
In contrast, when the temperature of the cold reservoir is smaller or of the order of the smoothness of the transmission probability,  $T_\text{c}\lesssim\gamma$, see the right column of Fig.~\ref{fig:DeltamuforMaxJc}, the cooling power is strongly suppressed and finite only if $\Delta T\ll T_\text{c}$. Its optimal value represents a \textit{local} maximum, see panel (b), which is reached in the linear response regime also with respect to the potential bias, $\Delta\mu\simeq\gamma$, as shown in panel~(f). The situation leading to this maximum is sketched in panel (d). The reason for the required linear-response scenario is that the barrier imposed by the smoothed step transmission function fails to be opaque against the tunneling of electrons from the hot reservoir even for a high values of $\dmu$ and $\Delta T$.

%%%%%%%%%%%%%%%%%%%%%%%%%%%%%%%%%%%%%%%%
  \begin{figure}[t]
    \centering
    \includegraphics[width=1\columnwidth]{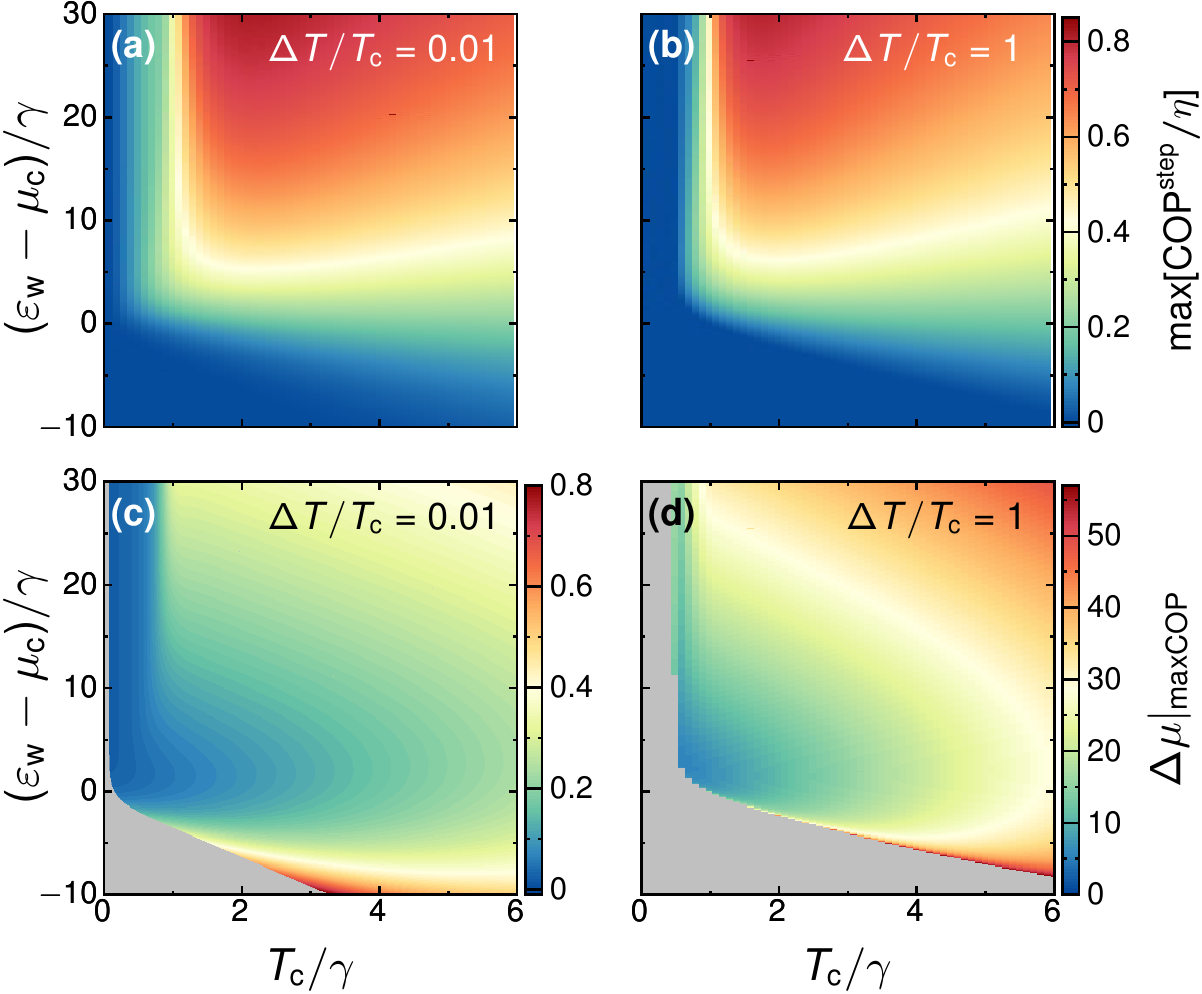}
    \caption{COP maximized with respect to $\Delta \mu$ for a conductor with smooth step transmission probabilities (a) for $\Delta T/T_\text{c}=0.01$ and (b) for  $\Delta T/T_\text{c}=1$. The corresponding potential biases $\mumaxCOPstep$, at which the COP becomes maximal are shown in (c) and (d).  
    \label{fig:MaxCOP_QPC}}
\end{figure}
%%%%%%%%%%%%%%%%%%%%%%%%%%%%%%%%%%%%%%%%

Finally, Fig.~\ref{fig:MaxCOP_QPC}~(a) and (b) show the maximum coefficient of performance, $\maxCOPstep$ and the potential bias, $\mumaxCOPstep$, at which it is reached. The maximum COP is particularly large in those regions in which the cooling power is non-zero but strongly suppressed. The reason for this is---as discussed before---that the particle current is small in this regions, thereby decreasing the absorbed power, $P$. 
The potential bias at which the maximum COP is reached has to be small enough in order to minimize the absorbed power and at the same time large enough to separate the hot and cold electronic distributions from each other in energy space, whenever the temperature bias is large. This leads to the fact that $\mumaxCOPstep<\mumaxJcstep$ and to the following features, which can be observed in panels~(c) and (d) of Fig.~\ref{fig:MaxCOP_QPC}. The magnitude of $\mumaxCOPstep$ strongly depends on the temperature bias; it is smaller than $\gamma$ for $\Delta T/T_\text{c}$ in the whole displayed parameter range of panel~(c), but quickly reaches values of tens of $\gamma$ when the temperature bias is increased to $\Delta T/T_\text{c}=1$. The potential bias needed to reach an optimal COP is in general small in those regions where the COP itself is small. An exception is the range of negative $\ewell-\mu_\text{c}$. In this case the step position is in between the electrochemical potentials of cold and hot reservoir. This requires a large bias to avoid a heat flow out of the hot reservoir.

%%%%%%%%%%%%%%%%%%%%%%%%%%%%%%%%%%%%%%%%%%%%%%%%%%%%%%%
\subsection{Conductor with  peaked transmission probability}\label{sec:peak}
%%%%%%%%%%%%%%%%%%%%%%%%%%%%%%%%%%%%%%%%%%%%%%%%%%%%%%%

We here discuss the cooling performance of a conductor with a Lorentzian-shaped peaked transmission probability,
see Eq.~(\ref{eq:peak}).
 We expect features of this peak-shaped transmission probability to occur in the cooling power of the QSH device, when the overall electrochemical potential is positioned in the vicinity of $\varepsilon_1=\varepsilon_\perp$ and all other energy scales imposed by the contacts, namely both temperatures and the potential bias are smaller than the distance between peaks. 

\begin{figure}[t]
    \centering
    \includegraphics[width=\columnwidth]{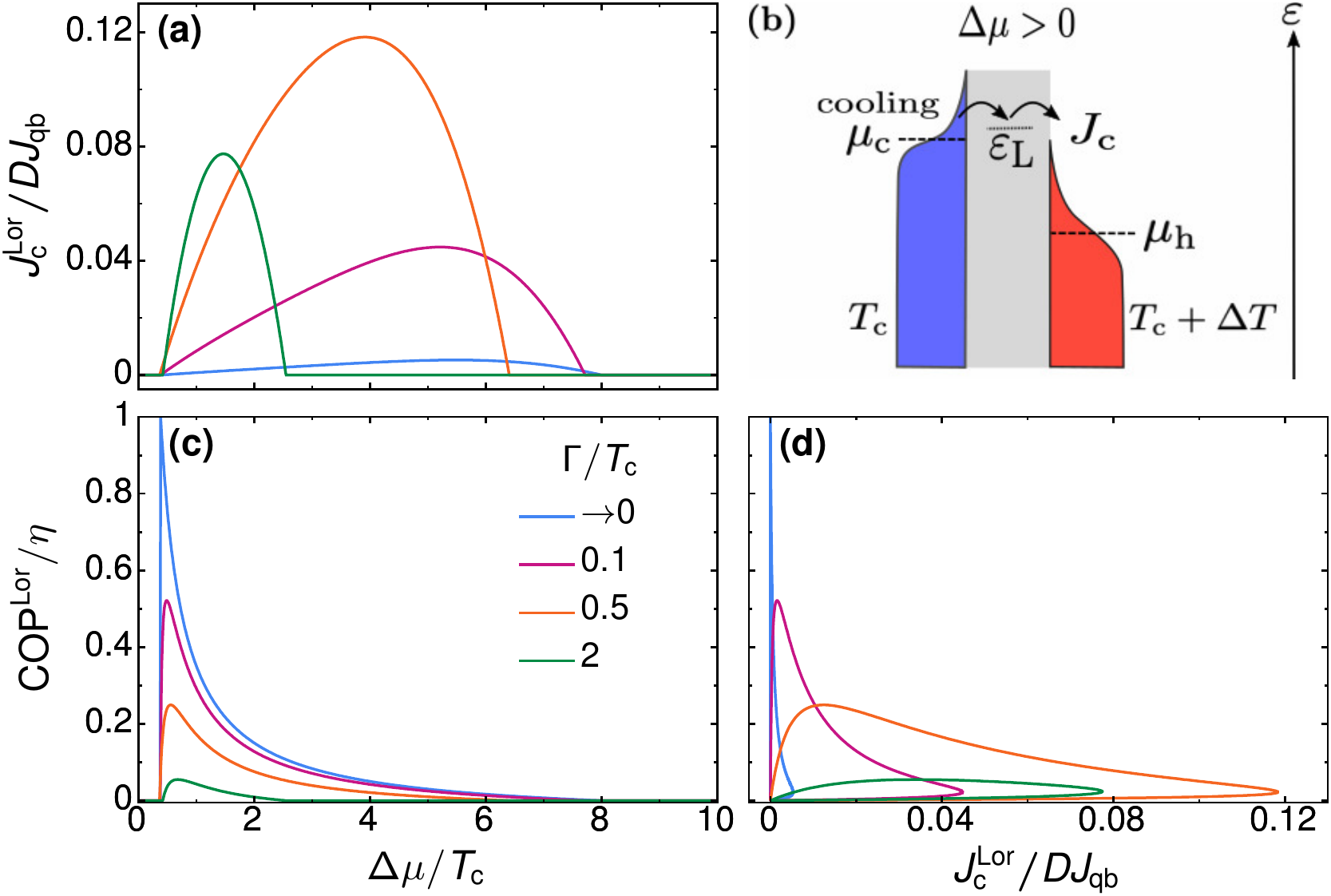}
    \caption{Cooling performance of the Lorentzian-shaped transmission probability for different broadening $\Gamma/T_\text{c}$ as indicated in the legend. We show (a) the cooling power and (c) the COP  as a function of $\Delta \mu/T_\text{c} $.  Lasso diagrams in panel~(d) show the COP at every cooling power while $\Delta\mu$ is varied.
     In all three panels (a), (c) and (d), the following parameters are fixed: $\mu=0$, $\varepsilon_\text{L}/T_\text{c}=4$, and $\Delta T/T_\text{c}=0.1$. Panel (b) shows the energy landscape for the Lorentzian-shaped transmission probability connecting a hot and cold electronic distribution. The displayed choice of potential allows the electron-like excitations to leave the cold  reservoir in a uni-directional manner.
    \label{fig:resonance_performance}}
\end{figure}

%%%%%%%%%%%%%%%%%%%%%%%%%%%%%%%%%%%%%%%%
%\begin{figure}[t]
 %   \centering
  %  \includegraphics[width=\columnwidth]{peak_sketch.png}
   % \caption{Energy landscape for a conductor with a peaked transmission probability %connecting a hot and cold electronic distribution. The displayed choice of %potentials allows the (a) electron-like excitations and (b) hole-like excitations %to leave the cold  reservoir in a uni-directional manner.}
    %\label{fig:peak_sketch}
%\end{figure}
%%%%%%%%%%%%%%%%%%%%%%%%%%%%%%%%%%%%%%%%
The principle of cooling via a peak-shaped transmission probability is demonstrated in Fig.~\ref{fig:resonance_performance}(b). The transmission peak serves as a precise energy-filter allowing transfer of electron- or hole-like excitations only in a sharp energy-window. 

We find the following analytical expression for the cooling power
\begin{align}\label{eq:Jcpeak}
    \Jcpeak=& \frac{\Dconst\Gamma}{4\pi}\text{Im}\left\{\left(-i\Gamma\right)\left(\Delta X+2\pi i\Delta f(i\varepsilon_\text{cutoff})\right)\right.\\
    &+\left.
    \left(\epeak-\mu_\text{c}-i\Gamma\right)
    \left(-\Delta Z+2\pi i \Delta f(\epeak-i\Gamma)\right)\right\}\nonumber\ ,
\end{align}
where we have introduced the functions  $\Delta Z=Z_\text{c}-Z_\text{h}$, and $\Delta X=X_\text{c}-X_\text{h}$ with $Z_\alpha=\tilde{\Psi}\left(\left[\epeak-\mu_\alpha-i\Gamma\right]/T_\alpha\right)$ and $X_\alpha=\tilde{\Psi}\left(\left[\mu_\alpha+i\varepsilon_\text{cutoff}\right]/T_\alpha\right)$. The expression $\tilde{\Psi}(x)=\Psi\left(\frac{1}{2}+\frac{ix}{2\pi}\right)+\Psi\left(\frac{1}{2}-\frac{ix}{2\pi}\right)$ contains the Digamma function $\Psi$. We have furthermore introduced the cutoff energy $\varepsilon_\text{cutoff}$, which represents the bandwidth of the contacts and is here chosen to be the largest energy scale. In this realistic case, in which we assume flat bands in the contacts, the result is independent of this cut-off energy.

In order to calculate the coefficient of performance, we again need to compare the cooling power, Eq.~(\ref{eq:Jcpeak}), to the absorbed power, which in the case of the peaked transmission probability is given by 
\begin{align}\label{eq:Ppeak}
    P^\text{Lor}=\Delta \mu \frac{\Dconst\Gamma}{4\pi}  \text{Im}\left\{
\left(-\Delta Z+2\pi i \Delta f(\epeak-i\Gamma)\right)\right\}\ .
\end{align}

%%%%%%%%%%%%%%%%%%%%%%%%%%%%%%%%%%%%%%%%
%%%%%%%%%%%%%%%%%%%%%%%%%%%%%%%%%%%%%%%%
Typically, the cooling power of a single transmission peak is small as long as the broadening is small, since it limits the energy-window in which excitations can be evacuated. This can be seen in Fig.~\ref{fig:resonance_performance}(a), showing that the maximum values of the cooling power are far below its quantum bound. At the same time, it has been found that ideal efficiencies can in principle be reached in the linear-response regime for peaked transmissions~\cite{Hicks1993May,Hicks1993Jun,Mahan1996Jul,Humphrey2002Aug}, due to the precise filtering and the small particle current, i.e. a small absorbed power, for the cooling process. 
%%%%%%%%%%%%%%%%%%%%%%%%%%%%%%%%%%%%%%%%
  \begin{figure}[t]
    \centering
    \includegraphics[width=\columnwidth]{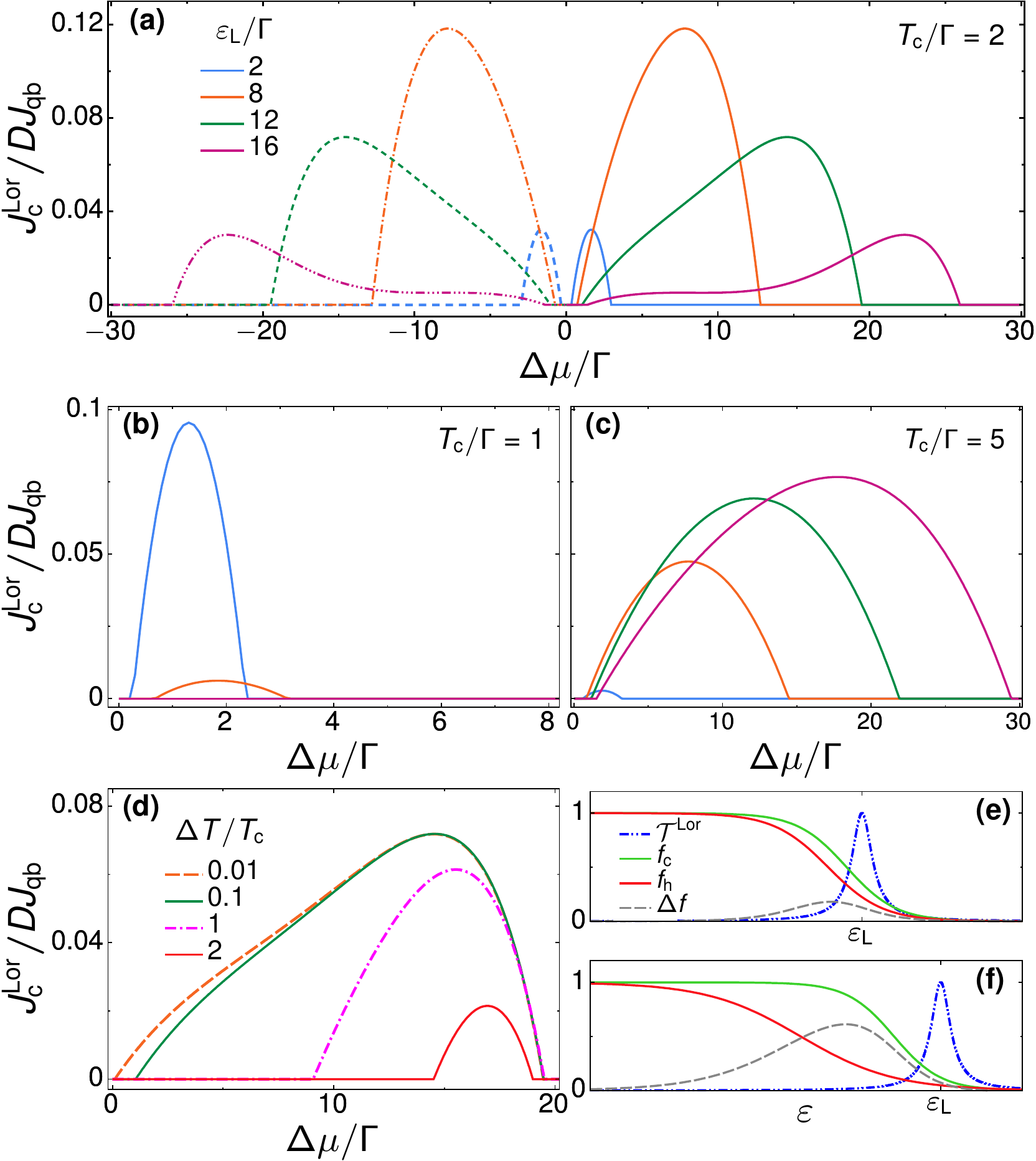}
    \caption{Cooling power as function of $\Delta \mu$ with $\mu=0$ of a conductor with Lorentzian-shaped transmission probability. In panels~(a) to (c) $\Delta T/T_\text{c}=0.1$ is fixed while different values for $\varepsilon_\text{L}$ are chosen (indicated in the legend) and the temperature regime is varied as (a) $T_\text{c}/\Gamma=2$ (b) $T_\text{c}/\Gamma=1$ and (c) $T_\text{c}/\Gamma=5$. In panel~(d) the cooling power at different temperature biases is shown while we fix $T_{\text{c}}/\Gamma=2$ and  $\varepsilon_\text{L}/\Gamma=12$. Panels (e) and (f) show transport window and the Lorentzian-shaped transmission probability for $T_{\text{c}}/\Gamma=2$. Other parameters are $\Delta T/T_{\text{c}}=0.1$, $(\varepsilon_\text{L}-\mu)/\Gamma=2$ and $\Delta\mu/\Gamma=1.5$ for panel~(e) and $\Delta T/T_{\text{c}}=1$, $(\varepsilon_\text{L}-\mu)/\Gamma=8$ and $\Delta\mu/\Gamma=8$ for panel~(f).
    \label{fig:resonance_Jc}}
\end{figure}
%%%%%%%%%%%%%%%%%%%%%%%%%%%%%%%%%%%%%%%%%
Also this characteristic is confirmed in Fig.~\ref{fig:resonance_performance}, where we show the COP of a device with a peaked transmission probability. Nonetheless, large COPs are reached only at low cooling powers, as shown by the lasso plots in panel~(d). Compared to the step-like transmission, where cooling turned out to be rather efficient even at large output powers, see Figs.~\ref{fig:Step}, lasso plots of the peaked transmission probability have contributions mostly in the vicinity to the plot axes.

Despite this limited performance of the peak-shaped transmission probability, its cooling power shows a number of interesting features, which can be identified also in the cooling power and COP of the QSH device introduced in Sec.~\ref{sec:model} and the rectangular barrier introduced in Appendix~\ref{app:RB}. In the following, we analyze the cooling power and COP obtained from Eqs.~(\ref{eq:Jcpeak}) and (\ref{eq:Ppeak}).

%%%%%%%%%%%%%%%%%%%%%%%%%%%%%%%%%%%%%%%%
  \begin{figure}[t]
    \centering
    \includegraphics[width=\columnwidth]{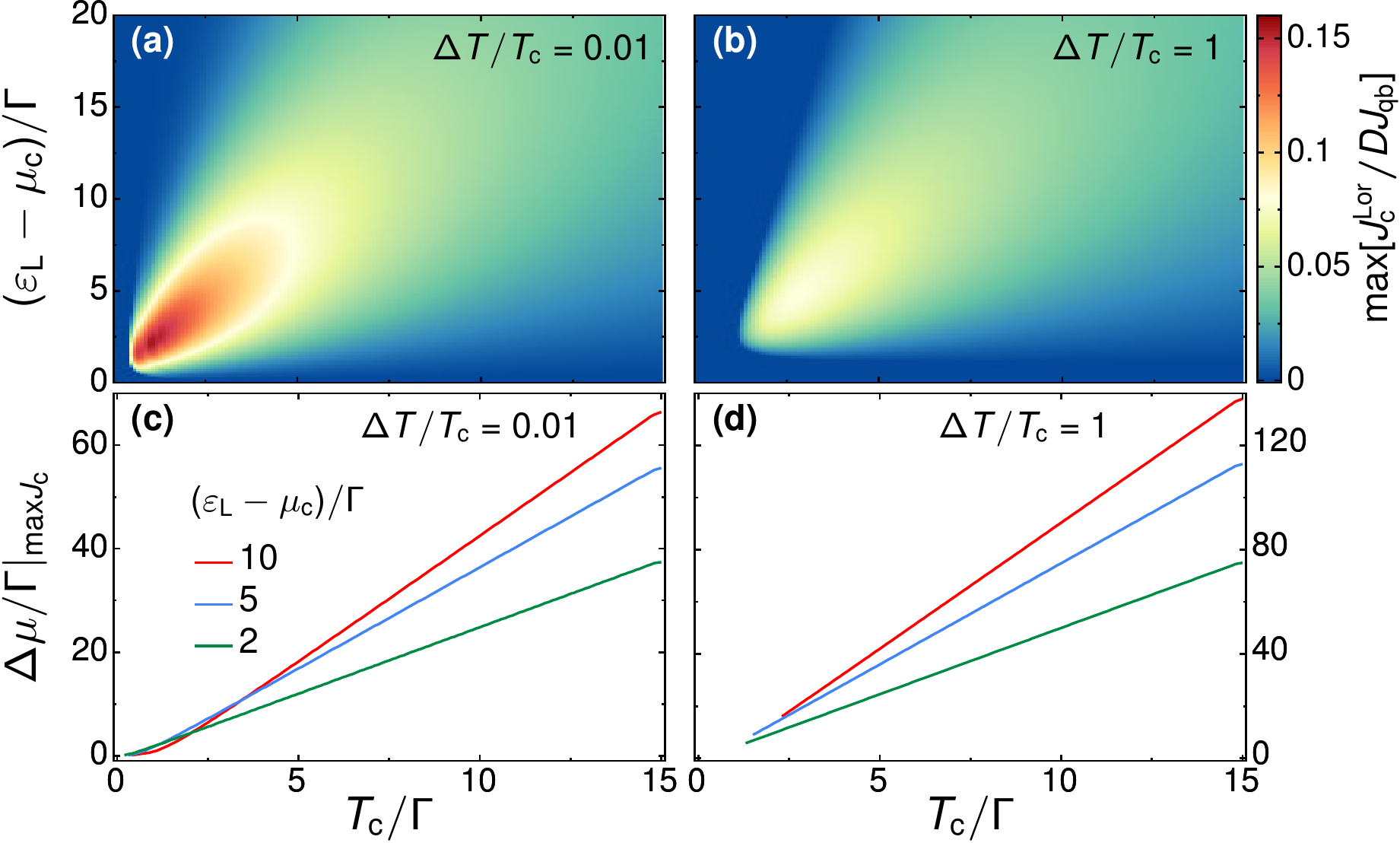}
    \caption{(a) Cooling power maximized with respect to $\Delta\mu$, $\maxJcpeak$, for a conductor with peaked transmission probability for $\Delta T/T_\text{c}=0.01$ and (b) for $\Delta T/T_\text{c}=0.01$.  The corresponding $\mumaxJcpeak$ at which $J_\text{c}$ is maximal is shown in panels~(c) and (d).
    \label{fig:MaxJc_Peak}}
\end{figure}
%%%%%%%%%%%%%%%%%%%%%%%%%%%%%%%%%%%%%%%%

The main properties of the cooling power can be seen from Fig.~\ref{fig:resonance_Jc}(a). Due to electron-hole symmetry, the cooling power is symmetric around $\mu=0$ under simultaneous exchange of $\varepsilon_\text{L}\rightarrow-\varepsilon_\text{L}$ and $\Delta\mu\rightarrow-\Delta\mu$, as expected from the sketch in Fig.~\ref{fig:CoolingPowerWell}(a) and (b). We now focus on the range of positive $\epeak-\mu$. Maximum cooling power is reached when $\mu_\text{c}$ is in the vicinity of, but still below $\varepsilon_\text{L}$. However, we find non-vanishing contributions to the cooling power in the whole range from small to large $\Delta\mu$. It depends on the temperatures and the level broadening when these contributions are most relevant. This can be seen from the remaining panels of Fig.~\ref{fig:resonance_Jc} as described in the following. 

For low temperatures, such that $T_\text{c}/\Gamma\lesssim1$, cooling is possible only in the linear-response regime of small $\Delta \mu/\Gamma$ and $\Delta T/T_\text{c}$, see panel~(b). The operational principle allowing for cooling in this regime is demonstrated in Fig.~\ref{fig:resonance_Jc}(e): The transmission peak allows for transport of a rather extended ratio of electronic excitations from the cold reservoir. At the same time, transport of hole-like excitations, which would lead to heating of the cold reservoir is not fully blocked by the energy-filter of the conductor, but partially compensated by the non vanishing distribution of the hot reservoir. The optimal point for cooling power in the linear-response, low $T_\text{c}$ regime, thus derives from an intricate interplay between transmission and occupations of both contacts.
In contrast, in the high-temperature regime, see panel~(c) for $T_\text{c}/\Gamma=5$, it is favorable to increase the transport window, requiring also a shift of $\varepsilon_\text{L}$ away from $\mu$. In order to completely separate the flow of electrons and holes from the cold distribution, it is then required that $\varepsilon_\text{L}$ is shifted further away from $\mu_\text{c}$, such that only the tail of the broadened, peak-shaped transmission is used for cooling. See panel~(f) of Fig.~\ref{fig:resonance_Jc} for a sketch of this situation.

While the large-temperature regime leads to a \textit{maximum} of the cooling power at increased potential biases, in the regime of large temperature biases $\Delta T/\Tc\gtrsim1$, cooling is even completely hindered in the linear-response regime of small potential biases $\Delta\mu$. This can be seen in Fig.~\ref{fig:resonance_Jc}(d). 

The occurrence of two different types of operational regimes, as indicated in panels~(e) and (f) of Fig.~\ref{fig:resonance_Jc} can lead to a plateau-peak or even double-peak shape of the cooling power as function of $\Delta\mu$, see the pink line in Fig.~\ref{fig:resonance_Jc}(a). This coexistence of the two regimes occurs when $T_c,\Delta$, and $\Gamma$ are of the same order of magnitude and at the same time $\varepsilon_\text{L}-\mu$ is much bigger than $\Gamma$, but still of moderate magnitude.

To complete the picture, we finally study the cooling power and COP maximized with respect to the applied potential bias $\Delta\mu$.
Fig.~\ref{fig:MaxJc_Peak} shows that the maximum cooling power obtained with a peak-shaped transmission function is always much smaller than its quantum bound, $J_\text{qb}/2$. Largest values for the cooling power are obtained when the temperature of the cold reservoir, $T_\text{c}$, and the broadening of the transmission peak, $\Gamma$, are of the same order of magnitude. At the same time the distance of the peak from the electrochemical potential of the cold reservoir, $\varepsilon_\text{L}-\mu_\text{c}$, should be of the order of $T_\text{c}$. Note that in order to obtain $J_\text{c}/\Dconst J_\text{qb}>0.1$, the device needs to be operated in the regime of small temperature and potential biases! When temperatures are large compared to the broadening, also $\varepsilon_\text{L}-\mu_\text{c}$ needs to increase to allow for cooling, however large potential biases are then required and the obtained cooling power is strongly suppressed.
With increasing temperature bias, the overall features of the cooling power are maintained, however larger $T_\text{c}/\Gamma$ and $\varepsilon_\text{L}-\mu_\text{c}$ are required and the obtained cooling power is smaller.

\begin{figure}[t]
    \centering
    \includegraphics[width=\columnwidth]{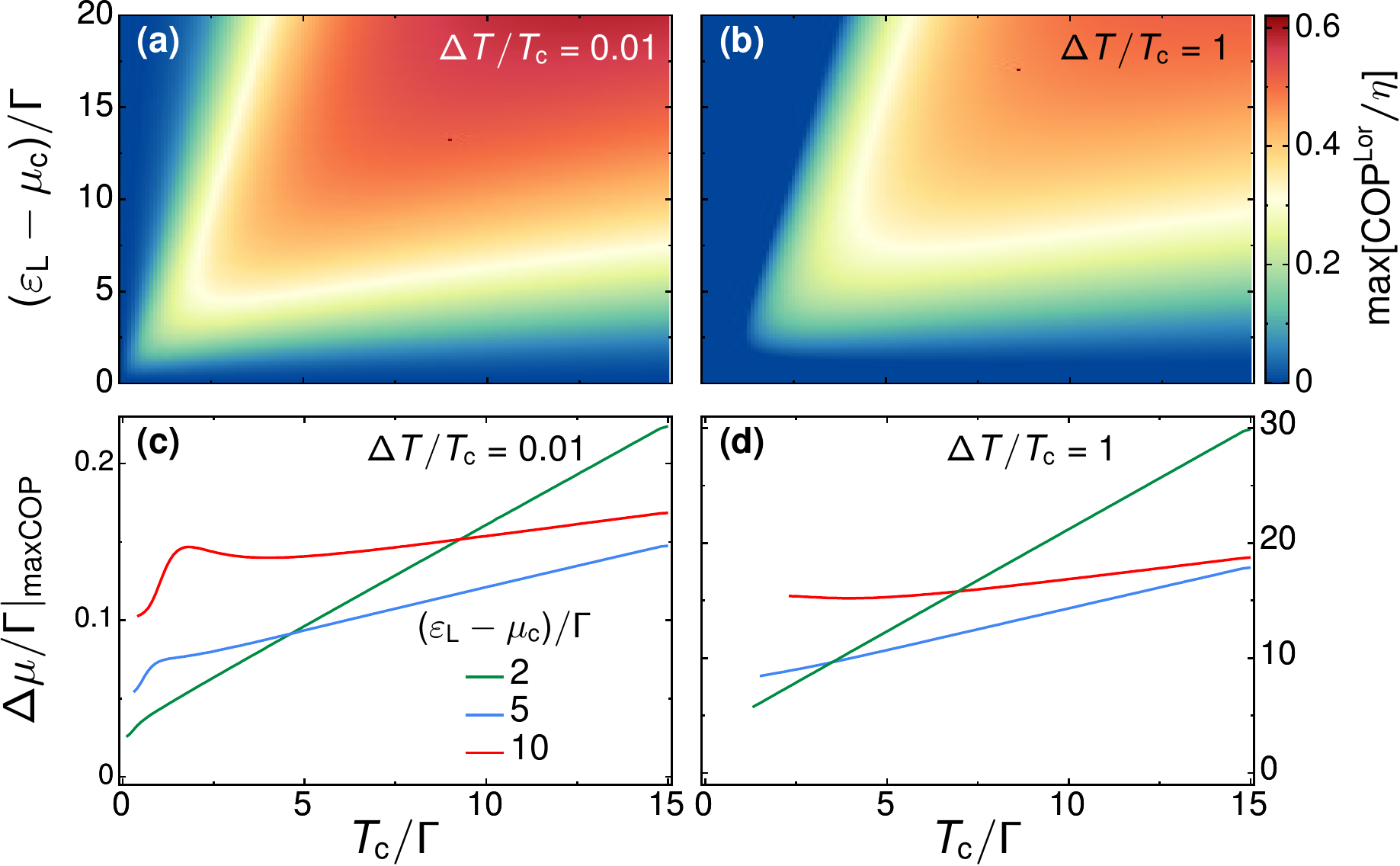}
    \caption{(a) COP maximized with respect to $\Delta\mu$, $\maxCOPpeak$ for a conductor with peaked transmission probability for (a) $\Delta T/T_\text{c}=0.01$ and (b) for $\Delta T/T_\text{c}=1$. The corresponding $\mumaxCOPpeak$ at which the COP is maximal is shown in panels~(c) and (d). \label{fig:MaxCOP_Peak}}
\end{figure}

Finally, the maximum coefficient of performance, shown in Fig.~\ref{fig:MaxCOP_Peak}, comes close to the Carnot value, when the broadening of the peak is the smallest energy scale. This is in contrast to the cooling power, which is maximal when $T_\text{c}$ and $\Gamma$ are of the same order. The voltages, required to maximize the COP, are again much smaller than those required to maximize the cooling power.

%%%%%%%%%%%%%%%%%%%%%%%%%%%%%%%%%%%%%%%%
%%%%%%%%%%%%%%%%%%%%%%%%%%%%%%%%%%%%%%%%%%%%%%%%%%%%%%%%%%%%%%%%%%%%%%%%%%%%%%%%
% \begin{figure}[t]
%    \centering
%    \includegraphics[width=\columnwidth]{transportwindowpeak.pdf}
%    \caption{Transport window and the Lorentzian-shaped transmission probability for %$\Gamma/T_{\text{c}}=0.5$. Other parameters are $\Delta T/T_{\text{c}}=0.1$, %$(\varepsilon_\text{L}-\mu_{\text{c}})/T_{\text{c}}=2$ and $\Delta\mu/T_{\text{c}}=2$ for %panel~(a) and $\Delta T/T_{\text{c}}=1$, $(\varepsilon_\text{L}-\mu_{\text{c}})/T_{\text{c}}=8$ %and $\Delta\mu/T_{\text{c}}=26$ for panel~(b).
%    \label{fig:transportwindowpeak}}
%\end{figure}
%%%%%%%%%%%%%%%%%%%%%%%%%%%%%%%%%%%%%%%

%%%%%%%%%%%%%%%%%%%%%%%%%%%%%%%%%%%%%%%%%%%%%%%%%%%%%%%
\subsection{QSH device}\label{sec:QSH}
%%%%%%%%%%%%%%%%%%%%%%%%%%%%%%%%%%%%%%%%%%%%%%%%%%%%%%%
With the preparation of the previous sections, we can now turn to analyze the nonlinear cooling performance of the QSH device. 
Actually, the transmission function $\TQSH$  contains features of all the conductors scrutinized previously. 
The main properties limiting its range of operation as a cooling device are the
gap, $2\varepsilon_{\perp}$, and the smoothness $\gamma$ for $\TQSH$ at the closing of the gap, as discussed in Sec. \ref{sec:range}. 
We recall that 
the smoothness is related to the width of the first peak as $\gamma=2\Gamma$, and is determined by the length $L/L_\perp$ of the magnetic island, as described by Eq.~(\ref{eq:tau}). How relevant these different features of $\TQSH$ are for the cooling performance, depends on their magnitude with respect to the temperatures and the potential bias.

%%%%%%%%%%%%%%%%%%%%%%%%%%%%%%%%%%%%%%%%%%%%%%%%%%%%%%%
\begin{figure}[t]
%\vspace{1.cm}
\begin{center}
\includegraphics[width=\columnwidth]{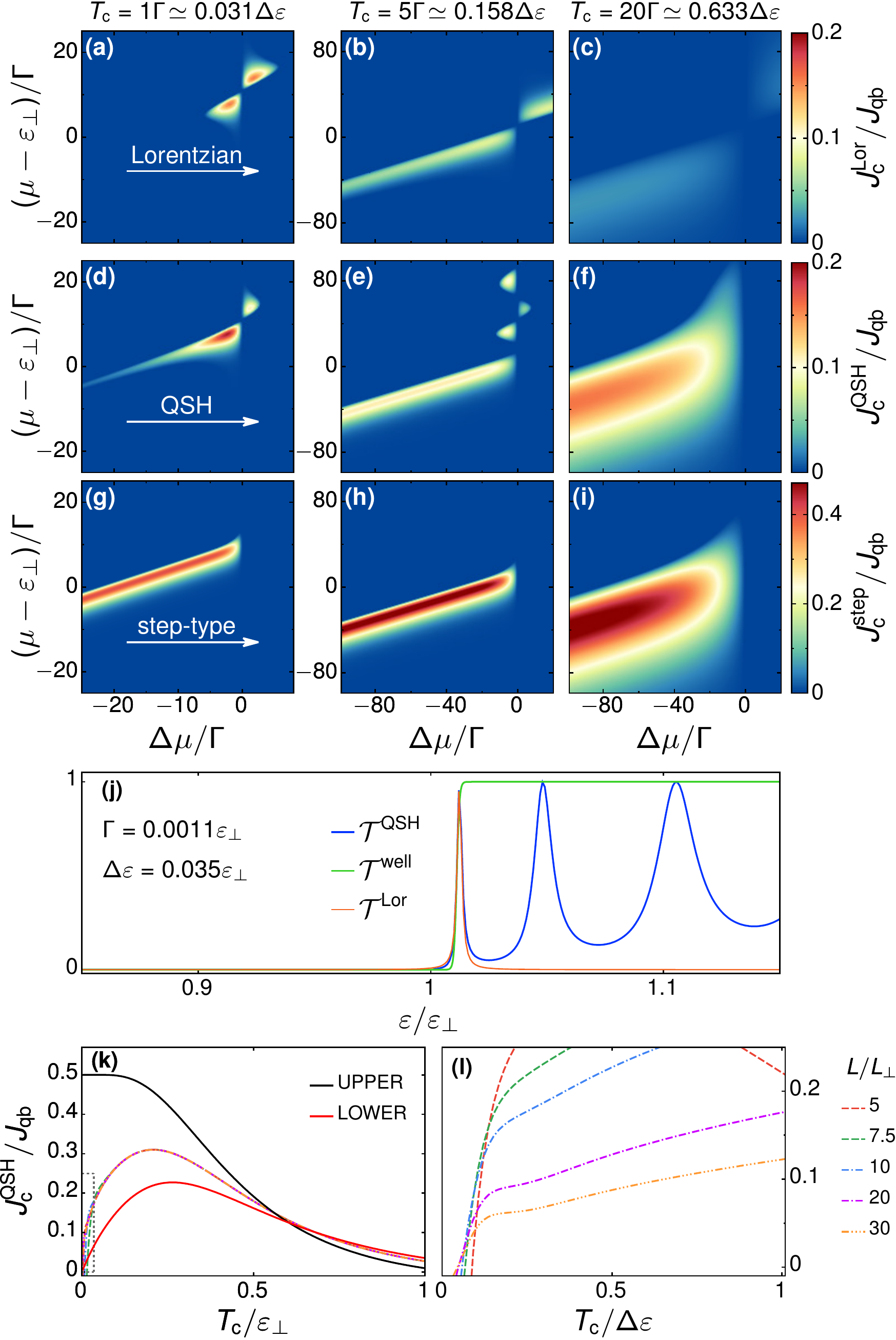}
\end{center}
\caption{Cooling power normalized by $J_{\rm qb}$ for $\Delta T/T_c=0.01$ of: (a-c) A conductor described by a Lorentzian transmission function of width $\Gamma$, (d-f) a
QSH device  with  $L/L_{\perp}=20 $, and (g-i) a conductor described by a step-type transmission function with broadening $\gamma=2 \Gamma$. (j) Sketch of the three transmission functions, showing how we fixed 
$\Gamma$ to be equal to the width of the first peak of the QSH device and $\gamma=2 \Gamma$ approximately equal to the slope at half the height of 
 $\TQSH$   at gap closing.  (k) Cooling power $J_{\rm c}^\text{QSH}$ as a function of $\Tc/\varepsilon_{\perp}$ for $\mu_{\rm c}=\varepsilon_{\perp}$, $\mu_{\rm h}=0.4\varepsilon_{\perp}$ and $\Delta T/T_c=0.01$. The calculations of the upper/lower bounds are, respectively calculated with $\Twell$ shown in (j) and the function $1-(\varepsilon/\varepsilon_{\perp})^2$, connecting the minimums. The plots for the different $L/L_{\perp}$ overlap in this scale.  (l) Zoom-in of the grey-dotted rectangular of panel (k), representing the scale of temperatures in units of the separation $\Delta \varepsilon$ between the first two peaks [see (j) for $L/L_{\perp}=20$ and text for other lengths].}
\label{fig:qsh-pow}
\end{figure}
%%%%%%%%%%%%%%%%%%%%%%%%%%%%%%%%%%%%%%%%%%%%%%%%%%%%%%%

In Fig.~\ref{fig:qsh-pow}, we show the evolution of the cooling power of a QSH device with $L/L_\perp=20$ as a function of $T_{\rm c}$ for a fixed ratio of $\Delta T/T_{\rm c}=0.01$ [see panels (d), (e) and (f)]. We compare these results with those
of a conductor described by a Lorentzian transmission probability of width $\Gamma$ [see panels (a), (b) and (c)], and also with a step-type transmission probability with $\gamma=2\Gamma$, corresponding to 
the envelope of the maximums of $\TQSH$ [see panels (g), (h) and (i)]. The  different transmission functions $\Tpeak$ and $\Twell$ are shown in Fig.~\ref{fig:qsh-pow}(j) along with $\TQSH$.
We see that for the lowest temperatures, $T_{\rm c}=\Gamma,\; 5 \Gamma$, the dominating feature is the structure of peaks of $\TQSH$. Hence, the behavior of the cooling power of the QSH device as a function of $\mu$ and $\Delta \mu$ resembles that of the conductor with Lorentzian transmission probability when we focus on the range $\mu\sim \varepsilon_{\perp}$ and small bias voltage $|\Delta \mu|/\Gamma \sim 1$. We can also clearly identify the
impact of the different peaks in the behavior of the cooling power for $\mu> \varepsilon_{\perp}$.  For large, negative $\Delta\mu$, the  faint feature resulting from the step-like feature in the transmission can already be seen. Instead, for higher temperatures, the transport window is broad enough to address several peaks at the same time and the behavior of $J_{\rm c}^\text{QSH}$ resembles that of
a step-like transmission function. 
Albeit, the magnitude is smaller in the case of the QSH device  by a factor of $\simeq 0.45$  with respect to the conductor modelled by the function $\Twell$ with unit height. 

More details on the evolution of $J_{\rm c}$ as a function of $T_{\rm c}$ and the role played by the different features are shown in panels (k) and (l). In this case, $\Delta T/T_{\rm c}$ is fixed at the same value as in the previous panels, while the chemical potentials are $\mu_{\rm c}=\varepsilon_{\perp}$ --just at the closing of the gap-- and $\mu_{\rm h}=0.4 \varepsilon_{\perp}$. 
Fig. ~\ref{fig:qsh-pow} (k) shows the results for different ratios $L/L_{\perp}$, along with the results corresponding to well-type transmission probabilities. The "upper" well-type transmission probability is defined as the envelope indicated in 
Fig. \ref{fig:qsh-pow} (j); the "lower" well-type transmission probability follows the minimums of the actual $\TQSH$, namely $1-(\varepsilon/\varepsilon_{\perp})^2$. These two envelopes therefore lead to  upper and lower bounds for the cooling power dominated by the step-type and well-type characteristic of the transmission probability of the QSH device. In the case of the upper bound, the parameters correspond to the regime of the well-type function illustrated in  Figs.~\ref{fig:CoolingPowerWell}(a) and (c), for which the upper quantum limit $J_{\rm qb}$ can be achieved at low enough temperatures. 
Results are shown for magnets of several lengths 
and the corresponding plots are almost indistinguishable one another within the scale of the figure. The most prominent feature of the universal behavior found at this scale 
is a maximum at $4 T_{\rm c} \simeq \varepsilon_{\perp} $ leading to a cooling power of approximately $60$ percent of the quantum bound. We have verified that this 
is consistent with a well-type transmission function defined by the envelopes of the maximums but with a height $D=0.61$. 

A zoom in the low-temperature regime
of Fig. ~\ref{fig:qsh-pow} (k) is shown in panel (l). In this case, the energy spacing between the first two peaks 
after the gap closing, $\Delta \varepsilon$, is a more appropriate scale to represent the temperatures
(see the horizontal axis). Here, we see that as the thermal broadening $\simeq 4  T_{\rm c}$  becomes larger than $\Delta \varepsilon $, the cooling power 
changes the slope as a function of $T_{\rm c}$ and, furthermore, it presents a shoulder-type behavior. We identify this change of behavior of $J_{\rm c}$ as a change from a thermoelectric response dominated by the peak after the gap closing of $\TQSH$ to one dominated by the step-type and well-type envelopes. This crossover becomes more pronounced for the largest values of $L/L_{\perp}$. Notice that
the value of $\Delta \varepsilon$, as well as the width of the peaks, depend on $L/L_{\perp}$, see Eq.~(\ref{eq:deltae}). For the lengths shown in the figure, these values are
$\Delta \varepsilon \simeq 0.4, 0.2, 0.13, 0.036, 0.016 \varepsilon_{\perp} $ for the cases  $L /L_{\perp}= 5, 7.5, 10, 20, 30$, with the corresponding $\Gamma$ being about
one order of magnitude smaller than $\Delta \varepsilon$ in each case. We see that, in order to properly capture this peak-to-step crossover, the low-energy scale $\Delta \varepsilon$ must be
sufficiently separated from the high energy scale defined by $\varepsilon_{\perp}$, which happens for large enough $L/L_{\perp}$.

%%%%%%%%%%%%%%%%%%%%%%%%%%%%%%%%%%%%%%%%%%%%%%%%%%%%%%%
\begin{figure}[t]
%\vspace{1.cm}
\begin{center}
\includegraphics[width=0.95\columnwidth]{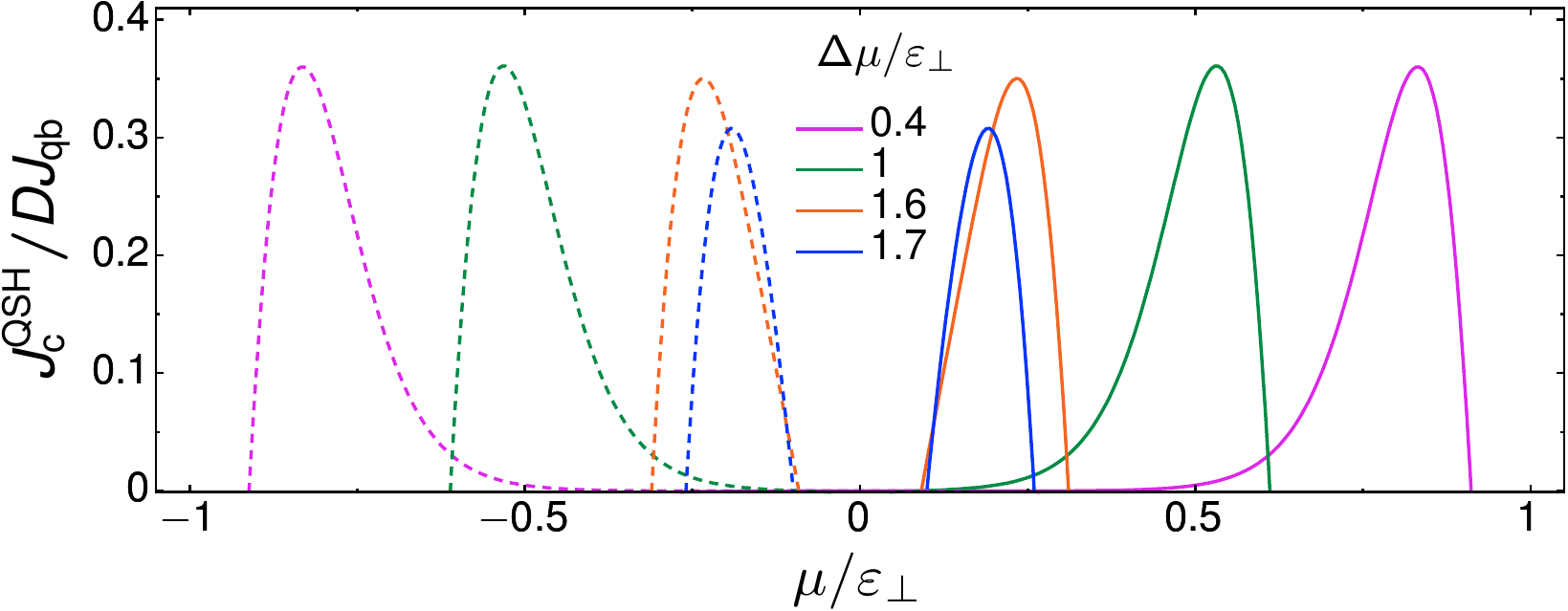}
\end{center}
\caption{Cooling power of the QSH device with  $L/L_{\perp}=20$, normalized with $D=0.61$. In order to compare with Fig.~\ref{fig:CoolingPowerWell}(c), we choose the other parameters equally, namely: $\Tc=0.05 \varepsilon_\perp$, $\Th=0.06 \varepsilon_\perp$.}
\label{fig:qsh-well}
\end{figure}
%%%%%%%%%%%%%%%%%%%%%%%%%%%%%%%%%%%%%%%%%%%%%%%%%%%%%%%

From the previous analysis, we conclude that below $T_{\rm c} \simeq \Delta \varepsilon/4$
the cooling response is equivalent to the one described by $\Tpeak$ with $\Gamma$ given by the width of the first peak and $D=1$. On the other hand, 
  for $T_{\rm c} > \Delta \varepsilon/4$ the cooling power of the QSH system  can be equivalently represented by an 
effective 
well-type
transmission function with $\gamma=2 \Gamma$ and height $D \simeq 0.61$ inferred from Fig.~\ref{fig:qsh-pow} (k).
This idea is further supported by Fig.~\ref{fig:qsh-well}, where we see that by normalizing with this effective height we can directly compare with Fig.~\ref{fig:CoolingPowerWell} for a purely well-type transmission probability. 

Having identified these different regimes, in which the cooling power of the QSH device resembles to the ones of a device with a step/well-shaped transmission or a peak-shaped transmission, is an important achievement. As a result, it is now possible to \textit{fully} transfer the analysis developed in the previous sections, Sec.~\ref{sec:well}-\ref{sec:peak}, to the analysis of the QSH device.

%The similarity between the QSH device and the effective description based on $\Tstep$ with height $D=0.615$ becomes apparent by comparing Fig.~\ref{fig:qsh-well} and Fig.~\ref{fig:CoolingPowerWell}(c). We see the equivalent behavior as a function of $\mu_{\rm c}$ at a qualitative and quantitative level. 

%%%%%%%%%%%%%%%%%%%%%%%%%%%%%%%%%%%%%%%%%%%%%%%%%%%%%%%
\begin{figure}[b]
%\vspace{1.cm}
\begin{center}
\includegraphics[width=\columnwidth]{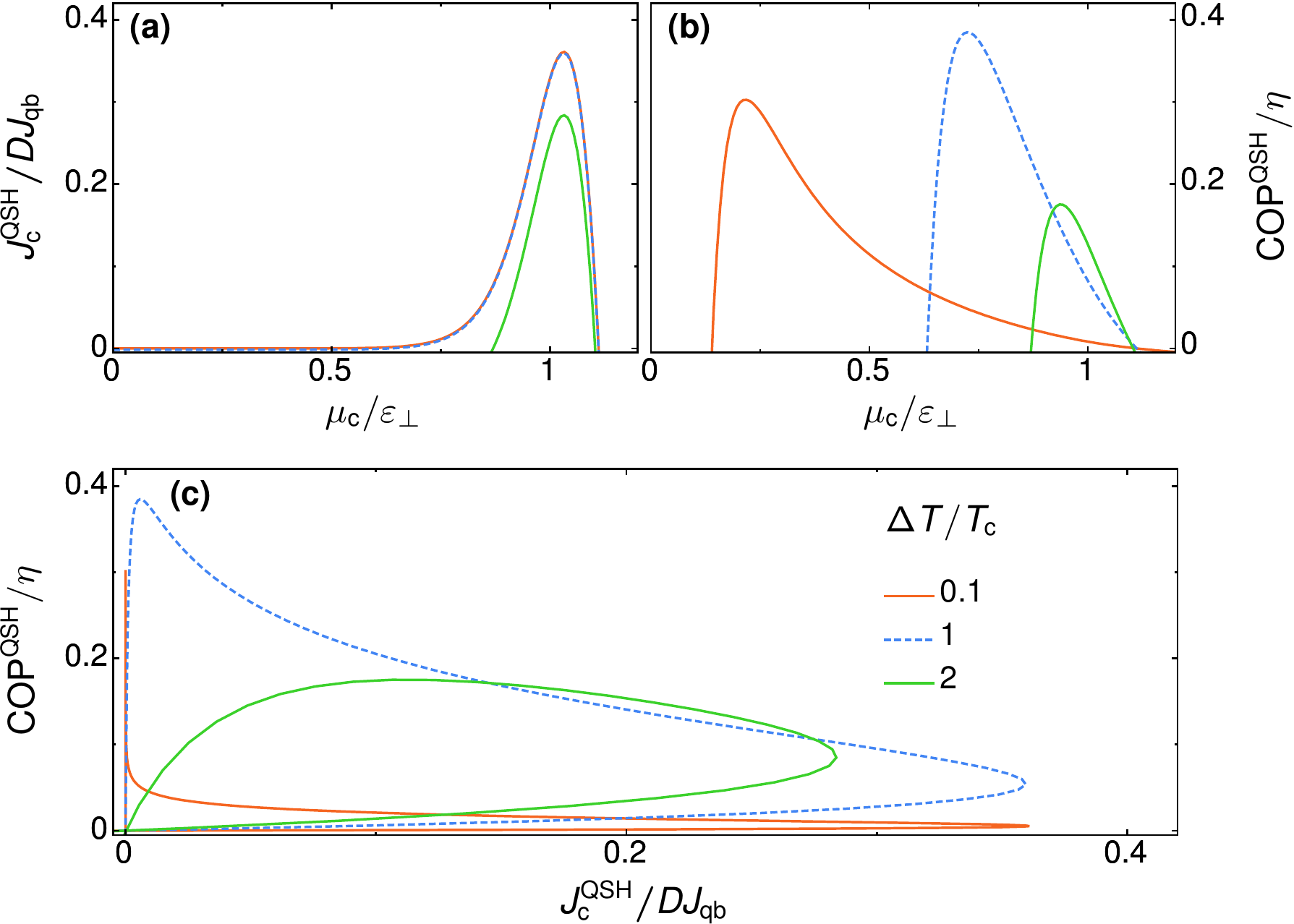}
\end{center}
\caption{Cooling performance for the QSH device with $L/L_{\perp}=20$, $\Tc=0.05 \varepsilon_\perp$ and $\muh=0$. (a) Cooling power, (b) COP and (c) lasso plot. The normalization factor is $D= 0.61$. 
}
\label{fig:qsh-performance}
\end{figure}
%%%%%%%%%%%%%%%%%%%%%%%%%%%%%%%%%%%%%%%%%%%%%%%%%%%%%%%

The analysis of the coefficient of performance along with the cooling power in the range of temperatures $T_{\rm c} > \Delta \varepsilon/4$, where the QSH device is expected to behave similarly to a system with a step/well-shaped transmission
is presented in Fig.~\ref{fig:qsh-performance}. Indeed, it is interesting to compare these figures with Fig.~\ref{fig:Step}, corresponding to $\Tstep$ for the same parameters
$\mu_{\rm c}, \mu_{\rm h}$, $T_{\rm c}$ and $T_{\rm h}$.  We see that the results for $J_{\rm c}$ are comparable for the two conductors, showing a similar behavior up to 
the normalization factor $D$ we discussed before. 
The behavior of the COP of the QSH device, shown in Fig. \ref{fig:qsh-performance} (b), is also comparable to the COP of a step-like transmission probability
  for the highest presented values
of 
$\Delta T/T_{\rm c}$. In these cases, the largest values of the COP are obtained for an operation where $\mu_{\rm c}$ is close to $\varepsilon_{\perp}$, as discussed in Sec.  ~\ref{sec:sharp}, where $\TQSH$ can be approximated by a step function.
Instead, for  the lowest value of $\dT/T_{\rm c}$ shown in the figure, the COP in the QSH device is lower than that of the conductor with a step-like transmission probability. 
The operational regime leading to the largest values of the COP in this case corresponds to $\mu_{\rm c}$ deep in the gap and, under these conditions, the well-type rather than the
step-type feature becomes dominant.
Nevertheless, in this regime, both
the cooling and electrical power are exponentially small and the COP is the result of the quotient between two vanishing small quantities.
 The simultaneous information of the two performance qualifiers is summarized in the lasso plot shown in the panel~(c) of Fig.~\ref{fig:qsh-performance}.

%%%%%%%%%%%%%%%%%%%%%%%%%%%%%%%%%%%%%%%%%%%%%%%%%%%%%%%
\begin{figure}[t]
%\vspace{1.cm}
\begin{center}
\includegraphics[width=\columnwidth]{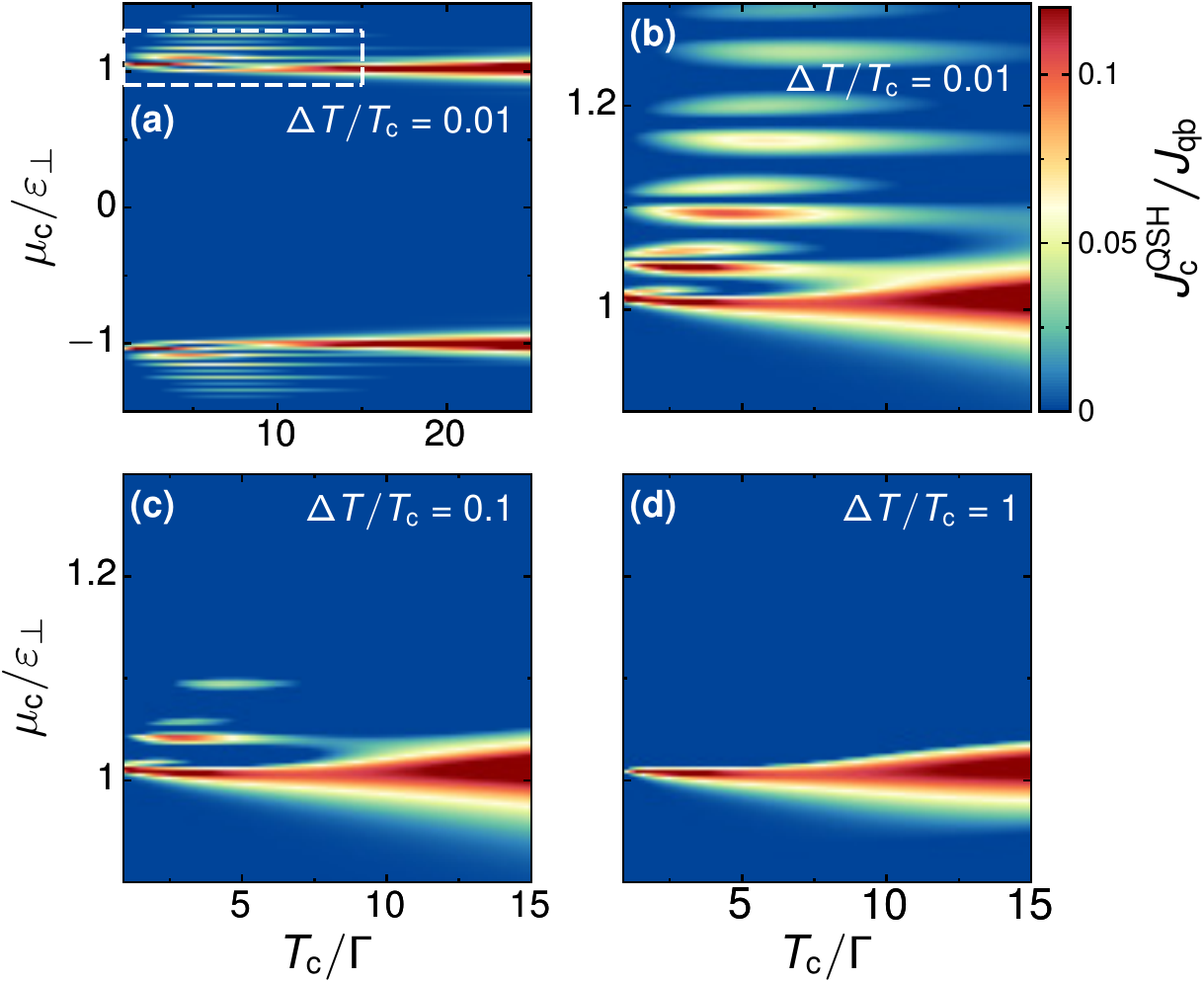}
\end{center}
\caption{Cooling power of the QSH device with  $L/L_{\perp}=20$, maximized over $\Delta \mu$ for different temperature biases. 
Panel (b) shows a zoom-in of panel (a) in the range indicated by the white-dashed rectangular. (c) and (d) show the same parameter range as in (b) but for larger temperature bias, $\dT/\Tc$.
}
\label{fig:qsh-max}
\end{figure}
%%%%%%%%%%%%%%%%%%%%%%%%%%%%%%%%%%%%%%%%%%%%%%%%%%%%%%%

The cooling power maximized over $\Delta \mu$, as a function of $\mu_{\rm c}$ and $T_{\rm c}$, for different different ratios of $\Delta T/T_{\rm c}$, is shown in Fig.~\ref{fig:qsh-max}. 
Here, we have chosen $\Gamma$, extracted as demonstrated in Fig.~\ref{fig:qsh-pow}, as the unit of energy, in order to be able to compare to Figs.~\ref{fig:MaxJc_QPC} and \ref{fig:MaxJc_Peak}. For this length of the magnet, we have the following relation between this scale and the separation between the first two peaks $\Delta \varepsilon \simeq 30 \Gamma$.
In panel~(a) a large range of values of $\mu_{\rm c}$ is shown, including negative values. The gap and the response generated by the sequence of peaks of $\TQSH$ after the closing of the gap are 
clearly distinguished. Panel~(b) shows a zoom in a range close to $\mu_{\rm c}\simeq\varepsilon_{\perp}$. For the lowest temperatures $\Tc$ and $\Th=\Tc+\dT$, shown in panels~(a)-(c), features of the peak-shaped transmission function are visible. In this regime the analysis presented in Sec.~\ref{sec:peak} applies. 
Panel~ (d) captures the regime of higher temperatures, where the structure of peaks becomes effectively interpolated, as discussed before, and the response becomes similar to that of a step function. In this case, the analysis presented in Secs.~\ref{sec:smooth} and \ref{sec:max} applies. In fact, notice, in particular, the similarity between the present figure and
Fig.~\ref{fig:MaxJc_QPC}.

\begin{figure}[tb]
%\vspace{1.cm}
\begin{center}
\includegraphics[width=\columnwidth]{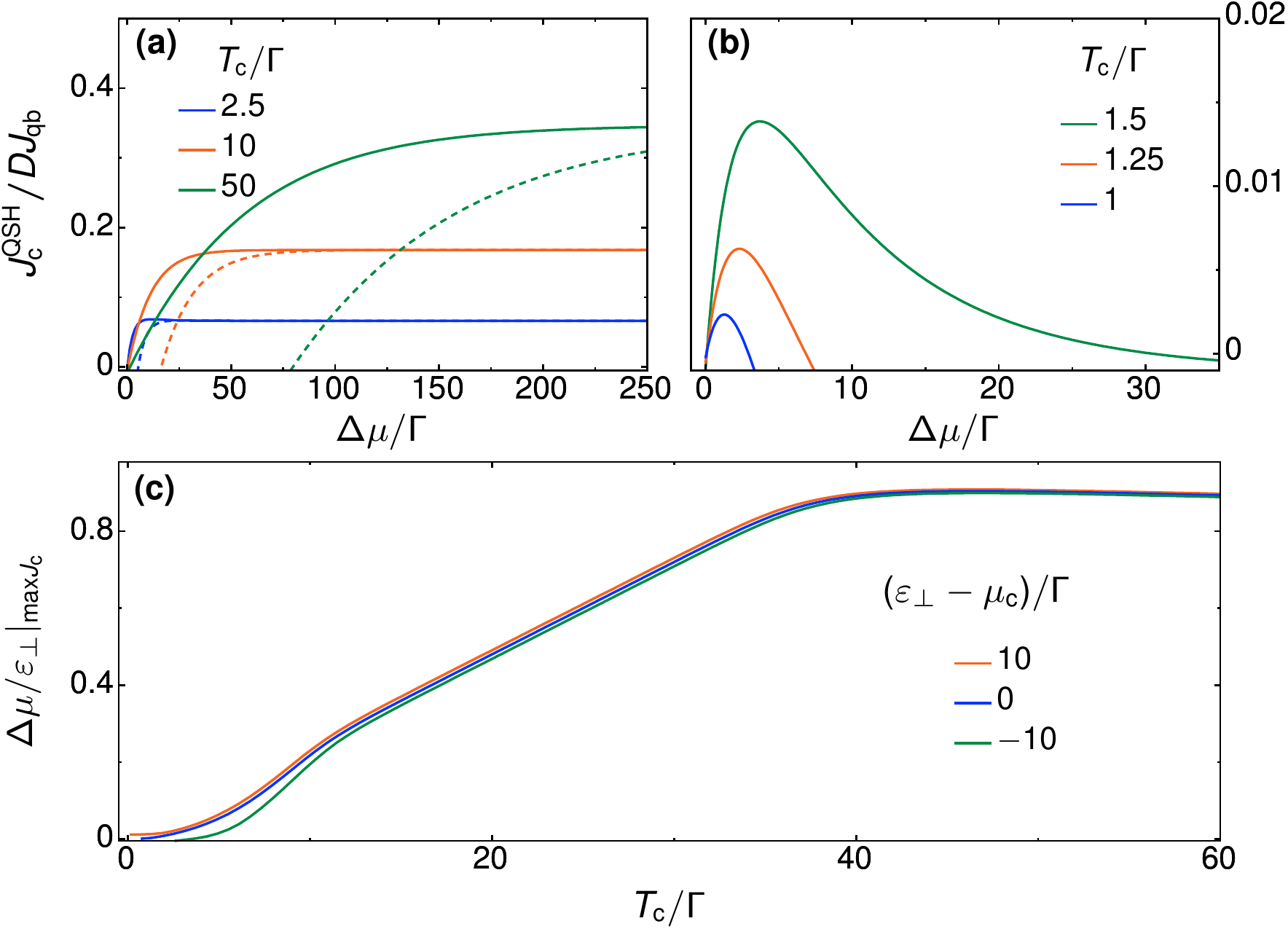}
\end{center}
\caption{Analysis of $\mumaxJcQSH$, where the cooling power of the QSH device is maximized. Solid and dashed lines correspond to $\Delta T/T_{\rm c}=0.1,\; 0.5$, respectively
for $L/L_\perp=20$ and $\mu_c=\varepsilon_{\perp}$. 
Panels (a) and (b): Cooling power as function of $\varepsilon_\perp-\muh\equiv\Delta\mu$; compare to Figs.~\ref{fig:DeltamuforMaxJc}(a)-(b) and Figs.~\ref{fig:MaxJc_Peak} (a)-(b).
Panel (c): $\mumaxJcQSH$ as function of $\Tc$  for the QSH device; compare to Fig.~\ref{fig:DeltamuforMaxJc}(e).  We distinguish three operational modes in the regimes $4 T_{\rm c}< \Delta \varepsilon$,
$\Delta \varepsilon < 4 T_{\rm c} < \varepsilon_{\perp}$ and $4 T_{\rm c}> \varepsilon_{\perp}$. For this length, $\Delta \varepsilon \simeq 30 \Gamma \simeq 0.036 \varepsilon_{\perp}$.
}
\label{fig:qsh-operation}
\end{figure}

Fig.~\ref{fig:qsh-operation} provides the complementary information to Fig. ~\ref{fig:qsh-max}. It shows $J_{\rm c}$ as a function of $\Delta \mu$ at selected values of $T_{\rm c}$ and
$\Delta T/T_{\rm c}$, with the electrochemical potential of the cold reservoir fixed at $\muc=\varepsilon_{\perp}$. The aim here is  to analyze which are the bias voltages that optimize the cooling power.
Panel~\ref{fig:qsh-operation}(a) focuses on the regime illustrated in Figs. ~\ref{fig:qsh-max}(c) and (d), where the QSH resembles a step-type conductor, while panel~\ref{fig:qsh-operation}(b) corresponds to the low-temperature
regime, where the peak-type behavior dominates. We clearly see that in the step-type regime the optimal operation corresponds to high voltages, as already discussed in 
Sec.~\ref{sec:max}, while in the case where the peak-feature dominates, the optimal operation is achieved for small $\Delta \mu$ and $\Delta T/T_{\rm c}$, as discussed in Sec.~ \ref{sec:peak}.
Finally, 
Fig.~\ref{fig:qsh-operation}(c) shows the bias $\Delta \mu$ that optimizes the cooling power as a function of the temperature of the cold reservoir. It is interesting to compare this figure 
with the corresponding ones for a smoothed step and Lorentzian functions, respectively,
Figs.~\ref{fig:DeltamuforMaxJc}(e) and  ~\ref{fig:MaxJc_Peak} (c). We clearly distinguish in
Fig.~\ref{fig:qsh-operation}(c) 
the low-temperature regime, $4T_{\rm c} < \Delta \varepsilon$ where
the behavior is dominated by the peak ,
corresponding to small bias voltages leading to the optimal operation, as in the case of  Fig. ~\ref{fig:MaxJc_Peak} (c). This is
followed by a
regime where $  \Delta \varepsilon < 4 T_{\rm c} < \varepsilon_{\perp}$, where the optimal operation corresponds to high voltages, as in the case of the step-like function
[see Fig.~\ref{fig:DeltamuforMaxJc}(e)]. At 
even higher temperatures, the well-type feature dominates.

%%%%%%%%%%%%%%%%%%%%%%%%%%%%%%%%%%%%%%%%%%%%%%%%%%%%%%%
\section{Summary and conclusions}\label{sec:conclusion}
%%%%%%%%%%%%%%%%%%%%%%%%%%%%%%%%%%%%%%%%%%%%%%%%%%%%%%%%%%
We have analyzed the nonlinear thermoelectric  cooling performance of a quantum spin Hall device, where backscattering is induced by means of a  magnetic island. The resulting transmission probability has a rich structure as function of energy: the dominating features are a well-type envelope due to the energy gap in the Dirac system introduced by the magnet, and oscillations above the closing of the gap. 
These features are manifest at different energy scales in the cooling performance of the device. This motivated an in-depth preliminary analysis of all of these features independently of one another. In our analysis, we have focused on the cooling power, but have also presented results on the coefficient of performance.

Importantly, this study is paramount also for the cooling performance of other conductors having transport properties determined by these features. In particular, to a quantum point contact of quantum Hall edge states, which is described by a step-like transmission function, and a quantum dot, which is described by a Lorentzian-type transmission function. 
%\textcolor{red}{Furthermore, the present study provides also a valuable guide for the understanding of the cooling properties of quantum Hall systems in the Corbino geometry, which can be accurately described in terms of transmission functions exhibiting some of these features\cite{corbino}. }
In a broad parameter regime for these cases we were able to provide analytical results. Beyond the present study, we expect them to also be useful for future studies of the different classes of devices addressed in this paper.
Based on this, we provide general insights into conditions for large cooling power and COP for these reference devices.
The transfer of these insights to the QSH device that motivated our work, allows us to provide an estimate of realistic parameters for future experiments here.

The main conclusions to highlight  for the operation of a future experiment using a QSH device for cooling are the following: (i) The well-type envelope of $\TQSH$, determined by the coupling between the magnetic island and the helical edges through the parameter $\varepsilon_{\perp}$, sets an upper limit to the operational temperatures for cooling. In contrast, the
smoothness of this envelope and the separation between peaks $\Delta \varepsilon$,  determined by the spatial extension of the magnet, $L/L_\perp$, set the lower-temperature properties.
Considering $\varepsilon_{\perp} \simeq 1-2 \times 10^{-4} e\text{V}$, the range of temperatures for which this system may operate as a cooling device is 
$T_{\rm c},\; T_{\rm h} \leq \varepsilon_{\perp} \sim 1.2 - 2.4$K.  Furthermore, considering this range of values for $\varepsilon_\perp$ together with 
$\hbar v_\text{F} \simeq 0.9$eV/nm as in the case of the HgTe 2D topological insulator~\cite{Koenig766,Roth294}, we find $L_{\perp} \sim 10 - 20 \mu$m~\cite{Gresta2019Oct}.
 (ii) Generally, the optimal cooling power is achieved if the chemical potential of the cold reservoir is placed in the vicinity of $\muc = \pm \varepsilon_{\perp}$, where the gap closes.
 The limit in the temperatures for which the effect of the peaks dominates the cooling response is set by $T_{\rm c}\sim \Delta \varepsilon /4$, which 
 corresponds to $T_{\rm c} \sim 10- 20$ mK for the case of a magnet of length $L=20 L_{\perp}$, and  $T_{\rm c} \sim 25- 50$ mK for a magnet of
 length $L=10 L_{\perp}$. In this low-temperature regime, the operation compares with that of a quantum dot device. 
Then, the cooling power is maximized in the linear-response regime, corresponding to low $\Delta \mu, \Delta T \ll \Gamma$, being $\Gamma$ the width of the first peak, which also depends on the
length of the magnet. Examples for realistic values are $\Gamma \sim 0.008 \varepsilon_{\perp}  \sim 10^{-7} -  10^{-6} e$V for $L/L_{\perp}=10,20$, with the mean chemical potential placed at half the width of the peak.
(iii) For higher temperatures, we enter a regime where the dominating feature is the \textit{step-like} envelope of $\TQSH$ and the behavior compares with that of a quantum point contact, modeled by a smooth step-type transmission function. 
In this regime, the optimal performance regarding the cooling power is
highly nonlinear in $\Delta \mu \sim \varepsilon_{\perp} \sim 10^{-4}e$V.  This is the regime of large cooling power, showing that a study of the nonlinear regime is crucial if one wants to get the device in good shape for future applications.

%%%%%%%%%%%%%%%%%%%%%%%%%%%%%%%%%%%%%%%%%%%%%%%%%%%%%%%
\section{Acknowledgements}
We are grateful for stimulating discussions with Matteo Acciai, Fabrizio Dolcini, Gwendal F\`eve, and Peter Samuelsson. We acknowledge financial support from CONICET, Argentina. We are sponsored by PIP-RD 20141216-4905 of CONICET,  PICT-2018-04536 and PICT-2017-2726 from Argentina, as well as  the Alexander von Humboldt Foundation, Germany (LA). Furthermore, we acknowledge financial support from the Knut and Alice Wallenberg foundation (N.D.,F.H.,J.S.) and from the Swedish Vetenskapsr\r{a}det (F.H, J.S.). We would also like to thank the hospitality of ICTP-Trieste (P.T.A., L.A.) under the support of Simons Foundation (L.A.) and the Federation ICTP-Institute (P.T.A.). 
%%%%%%%%%%%%%%%%%%%%%%%%%%%%%%%%%%%%%%%%%%%%%%%%%%%%%%%

%%%%%%%%%%%%%%%%%%%%%%%%%%%%%%%%%%%%%%%%%%%%%%%%%%%%%%%%%%%%%%%%
\appendix
%%%%%%%%%%%%%%%%%%%%%%%%%%%%%%%%%%%%%%%%%%%%%%%%%%%%%%%%%%%%%%%%

%%%%%%%%%%%%%%%%%%%%%%%%%%%%%%%%%%%%%%%%%%%%%%%%%%%%%%%%%%%%%%%%
\section{\label{app:RB} Mesoscopic conductor with extended, rectangular potential barrier}
%%%%%%%%%%%%%%%%%%%%%%%%%%%%%%%%%%%%%%%%%%%%%%%%%%%%%%%%%%%%%%%%

\begin{figure}[t]
%\begin{center}
\includegraphics[width=\columnwidth]{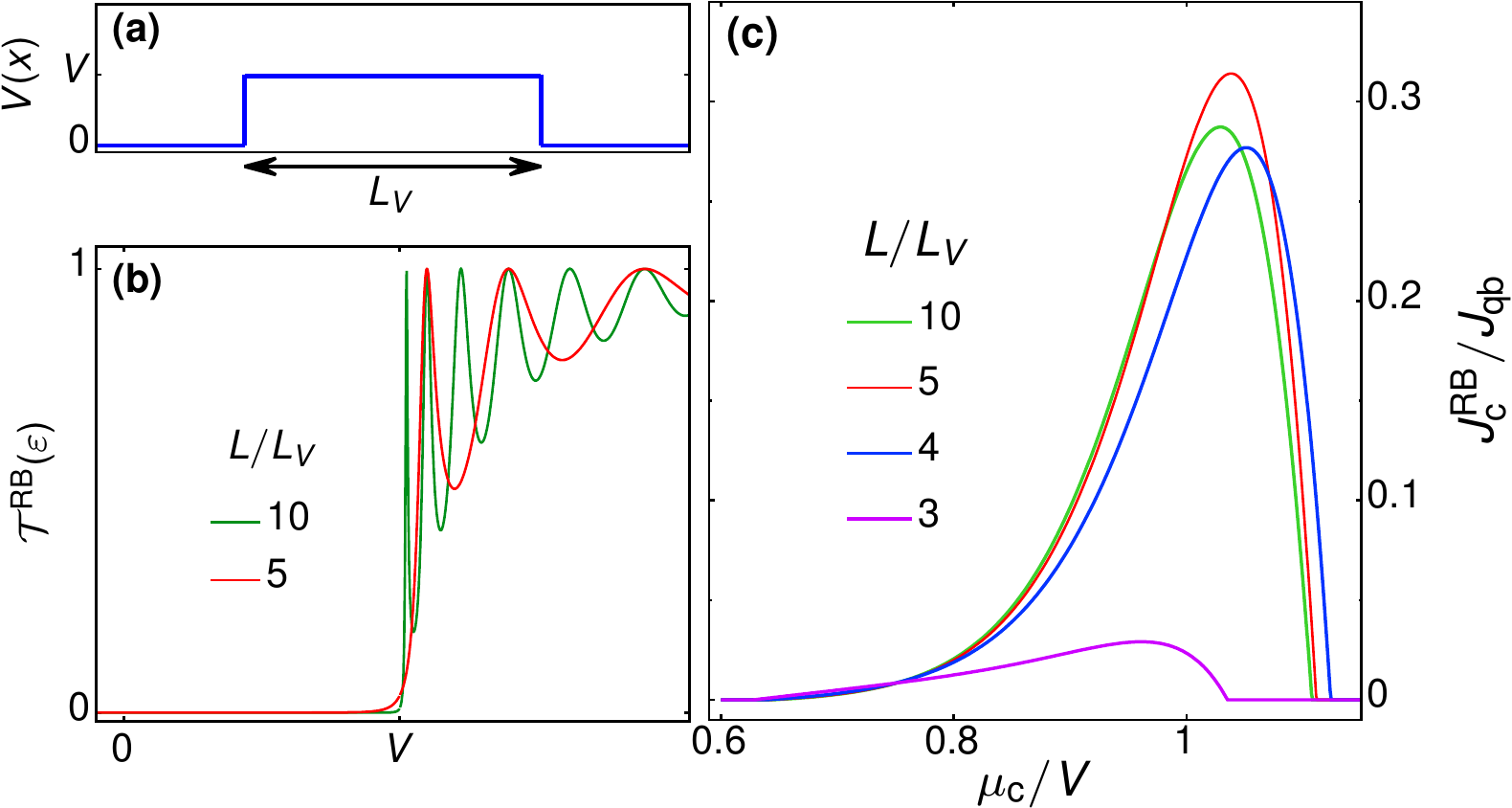}
%\includegraphics[scale=0.8]{App_MZI.pdf}
%\end{center}
\caption{(a) Potential landscape of a 1-dimensional conductor with a sharp, extended potential barrier, see Eq.~(\ref{appeq:RBpotential}). (b) Resulting  transmission probability, as given in Eq.~(\ref{eq:RB}) as a function of energy for different barrier lengths $L/L_V$. 
(c) Cooling power as function of $\mu_\text{c}$ for different values of $L/L_V$. The otherwise fixed parameters are $\mu_\text{h}/V=0.6$, $T_{\text{c}}/V=0.05$, and $\Delta T/T_\text{c}=0.1$.}
\label{appfig:RBmodel}
\end{figure}

In this appendix, we briefly present results for the cooling power of a coherent conductor with a potential barrier of height $V$ and spatial extension $L_V$
\begin{align}\label{appeq:RBpotential}
V(x)=\VRB\left[\Theta(x)-\Theta(x-L_V)\right].    
\end{align}
The potential landscape and the resulting transmission probability are shown in Fig.~\ref{appfig:RBmodel}(a) and (b). Strong similarities to the transmission probability of the QSH device presented in the main paper result from the backscattering induced by the extended potential region, leading to the step-like envelope function, and from the sharpness of the potential step, leading to oscillations. Note however that there are also important differences. These stem from the quadratic dispersion of the free electronic quasiparticles in the conductor at energies above the band bottom $E_\text{b}\equiv0$ (leading to a single step at \textit{positive} energies).

%%%%%%%%%%%%%%%%%%%%%%%%%%%%%%%%%%%%%%%%%%%%%%%%%%%%%%%%%%%%%%%%
%%%%%%%%%%%%%%%%%%%%%%%%%%%%%%%%%%%%%%%%%%%%%%%%%%%%%%%%%%%%%%%%

The analytic result for the transmission probability of the potential barrier given by Eq.~(\ref{appeq:RBpotential}) is known from standard textbooks on quantum mechanics and is given by
\begin{align}
\label{eq:RB}
\TRB(\varepsilon) = \frac{1}{1+ \dfrac{\VRB^2}{4\varepsilon|\varepsilon-\VRB|} \left|\sin(\lambda_V)\right|^2}. 
\end{align}
Here, we have introduced the complex, dimensionless parameter $\lambda_V=r_V(\varepsilon)L/L_V$ with the effective length $L_V=\sqrt{2\hbar^2/(m\VRB)}$ and the energy-dependent factor
\begin{align}
    r_V(\varepsilon)=\left\{\begin{array}{cc}
    2\sqrt{\varepsilon/\VRB-1}    \ , & \text{if} \ \ \varepsilon\geq \VRB \\
     2i\sqrt{1-\varepsilon/\VRB}  \ ,    & \text{if} \ \ \varepsilon< \VRB 
    \end{array}\right.
\end{align}
This allows a direct comparison to the expressions given in Eq.~(\ref{eq:tau}) for the QSH device with a magnetic region.

We show the cooling power of the conductor as a function of $\muc/\VRB$ in Fig.~\ref{appfig:RBmodel}(c) for different barrier lengths. There are clear similarities to analogous plots of the cooling power for a smooth step as shown in Fig.~\ref{fig:Step}(d). 
For sharp envelope functions, the cooling power has a maximum in the vicinity of $\muc/\VRB\approx1$. The suppression of the maximum value to lower values than $J_\text{qb}/2$ stems from the fact that the average value of finite transmission is reduced due to the oscillations. At small barrier lengths $L/L_V=3$, the envelope function is smooth, resulting in a reduced overall cooling power with a maximum shifted to smaller values of $\muc/\VRB$. 

Note that this trend can not be observed while decreasing $L/L_V$ from 10 to 4. The reason for this is that by decreasing the barrier length also the amount of oscillations entering into the temperature-broadened transport window changes.
A detailed analysis could be performed in analogy to the study of the QSH device in the main part of this manuscript, profiting from the insights obtained for conductors with well-, step-, and peak-shaped transmission probabilities. This analysis is however beyond the scope of this paper.

%%%%%%%%%%%%%%%%%%%%%%%%%%%%%%%%%%%%%%%%%%%%%%%%%%%%%%%%%%%%%%%%
\bibliography{cite.bib}
%%%%%%%%%%%%%%%%%%%%%%%%%%%%%%%%%%%%%%%%%%%%%%%%%%%%%%%%%%%%%%%%
\end{document}